\documentclass[a4paper,11pt]{article}
\pdfoutput=1 

\usepackage{jcappub}

\usepackage[T1]{fontenc} 
\usepackage{comment}

\DeclareUnicodeCharacter{2212}{\textminus}
\newcommand{\dis}[1]{\begin{equation}\begin{split}#1\end{split}\end{equation}}

\newcommand{\bfrac}[2]{\left(\frac{#1}{#2} \right)  }
\usepackage{xcolor,cancel}

\title{\boldmath A Possible Cosmological Origin of the KM3-230213A event}

\author[a,b]{Ki-Young Choi,}
\author[a]{Erdenebulgan Lkhagvadorj,}
\author[a,c]{and Satyabrata Mahapatra}

\affiliation[a]{Department of Physics and Institute of Basic Science, Sungkyunkwan University, 2066 Seobu-ro, Suwon-si, Gyeonggi-do, 16419, Korea}
\affiliation[b]{Korea Institute for Advanced Study, Seoul 02455, Korea}
\affiliation[c]{Indian Institute of Technology Goa,  Ponda-403401, Goa, India}

\emailAdd{kiyoungchoi@skku.edu}
\emailAdd{bulgaa@skku.edu}
\emailAdd{satyabrata@skku.edu}

\abstract{ We propose a novel cosmological scenario to explain the exceptional KM3-230213A neutrino event reported at an energy scale of $\mathcal{O}(100)$~PeV by the KM3NeT collaboration, along with its associated gravitational wave signatures. In our framework, ultra-high-energy neutrinos originate from the decay of a super-heavy sterile neutrino produced via the Hawking evaporation of primordial black holes in the early Universe. 
{While two sterile neutrinos in the model are responsible for generating light neutrino masses as required by oscillation data through the type-I seesaw mechanism, a third sterile neutrino or a heavy neutrino-like fermion with extremely feeble couplings can have a lifetime suitable for producing a neutrino flux consistent with the observed event.}
Furthermore, our scenario predicts two distinct GW signatures: one arising from gravitons emitted during PBH evaporation and another from the Bremsstrahlung process during the decay of the sterile neutrino. These complementary signals provide a multi-messenger probe of the underlying physics. Our results thus offer a compelling explanation for the KM3-230213A event and open new avenues for investigating the interplay between ultra-high-energy neutrino astronomy and ultra-high-frequency gravitational wave cosmology.
 }

\hypersetup{
colorlinks = true,
linkcolor = blue,
citecolor = magenta
}

\begin{document}
\maketitle
\flushbottom

\section{Introduction}\label{intro}

The deep-sea neutrino telescope KM3NeT recently reported the detection of an exceptionally high-energy neutrino event, KM3-230213A, with an energy of $\mathcal{O}(100)~{\rm PeV}$~\cite{KM3NeT:2025npi}. On 13 February 2023, the ARCA detector identified an ultra-high-energy muon event with an energy of $120^{+110}_{-60}~{\rm PeV}$; this measured muon energy provides a lower limit on the energy of the incoming neutrino. Based on simulations of the ARCA detector, the median neutrino energy required to produce such muons is estimated to be 220\,PeV, with 68\% (90\%) of simulated events falling in the range of 110–790\,PeV (72\,PeV–2.6\,EeV). Assuming an $E^{-2}_\nu$ spectrum, the corresponding neutrino flux is measured as 
$
E^2_\nu\Phi_\nu(E)=5.8^{+10.1}_{-3.7}\times 10^{-8}\,{\rm GeV\,cm^{-2}\,s^{-1}\,sr^{-1}},
$
as reported in~\cite{KM3NeT:2025npi}.

This event represents the highest-energy neutrino ever observed, exceeding the most energetic neutrino detected by IceCube by two orders of magnitude. Moreover, the absence of any conclusive astrophysical source in the KM3NeT analysis~\cite{KM3NeT:2025bxl,KM3NeT:2025aps} suggests a cosmic origin. While ultra-high-energy (UHE) cosmogenic neutrinos are expected from interactions between UHE cosmic rays and cosmic microwave background photons, the flux implied by the KM3NeT observation is in tension with standard cosmogenic neutrino predictions derived from data collected by the Pierre Auger Observatory and IceCube~\cite{KM3NeT:2025vut}. Furthermore, the non-observation of similar events by IceCube and the Pierre Auger Observatory introduces a $2.5\sigma$–$3\sigma$ tension with the cosmogenic origin hypothesis~\cite{KM3NeT:2025ccp}. Consequently, the KM3-230213A event has sparked considerable interest in exploring alternative explanations and potential new physics interpretations~\cite{Li:2025tqf,Fang:2025nzg,Satunin:2025uui,Dzhatdoev:2025sdi,Neronov:2025jfj,Amelino-Camelia:2025lqn,Yang:2025kfr,Boccia:2025hpm,Borah:2025igh,Brdar:2025azm,Kohri:2025bsn,Narita:2025udw,Wang:2025lgn,Alves:2025xul,Jiang:2025blz,Crnogorcevic:2025vou,Jho:2025gaf,Klipfel:2025jql, Dvali:2025ktz}. 
It is worth mentioning here that, the origin of PeV neutrino events at IceCube~\cite{IceCube:2013cdw, IceCube:2013low} reported more
than a decade ago, was also explored in several works including~\cite{Feldstein:2013kka,Rott:2014kfa,Boucenna:2015tra,Barman:2025bir}.

In this work, we explore a possible cosmological origin for the KM3-230213A event by considering the decay of a super-heavy particle produced in the early Universe via Hawking evaporation of primordial black holes (PBHs). Even if the particle’s mass approaches the Planck scale, it can be generated during the final stages of PBH evaporation, when the black hole temperature becomes extremely high. Once produced, this particle may decay into neutrinos with ultra-high energies. These neutrinos subsequently redshift to lower energies by the present epoch, potentially giving rise to the observed event.

{The $3\sigma$ confidence level (CL) flux range reported by KM3NeT~\cite{KM3NeT:2025npi} extends up to an order of magnitude below the exclusion limits from IceCube and Auger~\cite{ IceCube:2025ezc, AbdulHalim:2023SN}. It has been shown that, with diffuse isotropic flux assumptions, such as those arising from cosmogenic sources or relic particle decays, the tension reaches the $3.1\sigma$–$3.6\sigma$ level, with point-source or transient origin scenarios yielding up to $2$–$2.9\sigma$ inconsistency~\cite{Li:2025tqf}. In fact, the KM3NeT collaboration itself reported that a $2.2\sigma$ upward fluctuation would be needed to reconcile this event rate with IceCube's null observation~\cite{KM3NeT:2025npi}.
Our model predicts an isotropic neutrino flux 
as we consider a population of heavy right-handed neutrinos (RHNs) produced via Hawking evaporation of PBHs, which decay after the epoch of neutrino decoupling. This results in a cosmological and isotropic flux of UHE neutrinos in the present Universe. 
Given the large uncertainty in the KM3NeT event's flux, our model remains consistent with existing null results from IceCube and Auger, provided the actual flux lies near the lower end of the KM3NeT $2$–$3\sigma$ band. 
However, if one were to fit the central KM3NeT flux value, one inevitably re-encounters the $\sim3\sigma$ tension under diffuse flux hypotheses~\cite{Li:2025tqf}. In our parameter-space scan, we select benchmark points that lie below the IceCube’s upper limit while still within KM3NeT’s error-bar to explain the observed event. 
Nevertheless, this scenario, like others of similar cosmogenic origin, is subject to the same discrepancy, which may indicate either a rare upward fluctuation or a hint of physics beyond the Standard Model. 
Future data from KM3NeT~\cite{KM3Net:2016zxf}, IceCube-Gen2~\cite{IceCube-Gen2:2020qha}, and GRAND~\cite{GRAND:2018iaj} will help to clarify the nature of this intriguing event.}

{For a concrete framework, we adopt the type-I seesaw mechanism for generating light neutrino masses, extending the Standard Model with three heavy sterile neutrinos~\cite{Minkowski:1977sc,Gell-Mann:1979vob,Mohapatra:1979ia,Schechter:1980gr}. The relevant Lagrangian is 
$
\mathcal{L_\nu} \supset Y_\nu\, \overline{L}\,\tilde{H}\,N + \frac{1}{2} M_N\, \overline{N^c}\,N + \text{h.c.},
$
which, after electroweak symmetry breaking, yields  Majorana light neutrino masses 
$
M_\nu = {Y_\nu^2\,\langle H\rangle^2}/{M_N}$.
While two of the sterile neutrinos ($N_2$, $N_3$) have sizable couplings and are responsible for the observed neutrino masses via the seesaw mechanism, the third sterile neutrino, $N_1$, does not contribute appreciably to the light neutrino masses due to its extremely suppressed Yukawa coupling. Instead, it plays a distinct cosmological role in our framework.
Given the extremely high energy of the KM3-230213A event, we consider $N_1$ to be super-heavy, even up to the Planck scale so that the neutrino produced in its decay attains an enormous energy that redshifts to observable levels today. Since it is super heavy and its coupling to the Standard Model is exceedingly small, $N_1$ cannot be thermally produced in the early Universe. Instead, we propose that PBHs serve as the production mechanism via Hawking evaporation. The feeble coupling of $N_1$, allowed by the freedom in choosing the lightest active neutrino mass to be vanishingly small, enables its lifetime to be tuned appropriately to generate the required ultra-high-energy neutrino flux at a specific epoch in the early Universe, without conflicting with neutrino oscillation data. The presence of three RHNs is naturally motivated in scenarios like gauged $B-L$ model, where anomaly cancellation requires the inclusion of three SM singlet fermions.}
It is worth mentioning that, within the type-I seesaw framework for neutrino mass generation, the decays of the RHNs can also play a crucial role in explaining the baryon asymmetry of the Universe through the leptogenesis mechanism. However, in this setup, the conventional $N_1$-dominated leptogenesis is not viable. This is because $N_1$ must be long-lived and decay after neutrino decoupling to generate the ultra-high-energy neutrino flux required to explain the KM3NeT event. Owing to its extremely small coupling, the CP asymmetry parameter $\varepsilon_1$ is also negligibly small. As a result, although $N_1$ decays after neutrino decoupling, the generated lepton asymmetry is insignificant and has no substantial effect. Consequently, in our framework, we focus on leptogenesis driven by the decays of $N_2$ and $N_3$. {In our setup, we choose $N_1$ to be the heaviest of the three RHNs, followed by $N_3$ and then $N_2$. The RHNs $N_2$ and $N_3$, which have sizable Yukawa couplings, are responsible for generating the observed light neutrino masses and play an active role in generating the baryon asymmetry via leptogenesis. This mass ordering and role separation are made consistent with neutrino oscillation data through an appropriate choice of the orthogonal matrix in the Casas--Ibarra parametrization~\cite{Casas:2001sr}.
}

The reported arrival direction of KM3-230213A, nearly opposite to the Galactic Center, presents a particular challenge for conventional dark matter interpretations. Typical decaying dark matter scenarios, where signals originate from the galactic halo, would predict enhanced flux toward the Galactic Center due to higher DM density. This directional anomaly thus favors an extragalactic source. Our cosmological PBH-driven mechanism naturally accounts for this through its isotropic neutrino flux originating in the early Universe. Furthermore, although sterile neutrinos decay around Big Bang Nucleosynthesis (BBN), their energy density remains low enough to avoid disrupting light element abundances, satisfying constraints on late-decaying particles~\cite{Kawasaki:2017bqm,Yeh:2024ors}.

Our scenario also predicts two intriguing high-frequency gravitational wave (GW) signatures that are theoretically correlated with the neutrino flux required to explain the KM3-230213A event. The first arises from gravitons emitted during the Hawking evaporation of primordial black holes~\cite{Dolgov:2000ht,Anantua:2008am,Dolgov:2011cq,Dong:2015yjs,Arbey:2021ysg,Ireland:2023avg,Ireland:2023zrd}, which simultaneously generate the heavy sterile neutrino. Since the number and energy density of the sterile neutrinos and consequently the resulting ultra-high-energy neutrino flux depend on both the PBH energy density and the initial PBH mass, this GW component is theoretically linked to the same cosmological origin.
The second GW component arises from graviton emission via the Bremsstrahlung process during the decay of the sterile neutrino~\cite{Nakayama:2018ptw,Huang:2019lgd,Barman:2023ymn,Bernal:2023wus,Barman:2023rpg,Kanemura:2023pnv,Hu:2024awd,Choi:2024acs,Datta:2024tne,Inui:2024wgj,Jiang:2024akb}. Although graviton production in such decays is suppressed by the Planck mass (with the effective interaction given by $\mathcal{L}_{\rm eff} \supset \lambda\, h_{\mu\nu}\,T^{\mu\nu}$, where $\lambda\sim M_p^{-1}$, $h_{\mu\nu}$ is the graviton field, and $T^{\mu\nu}$ is the energy-momentum tensor), this suppression is compensated if the mother particle is extremely heavy, leading to a sufficient energy transfer to gravitons and high-frequency GW signatures.
{While these GW signals provide a theoretically interesting multi-messenger connection between the early Universe physics and the KM3-230213A event, we note that they lie in an ultra high frequency range that is currently far beyond the reach of existing or planned GW detectors. Nevertheless, the possibility of future developments in high frequency GW detection strategies may open new directions for probing such early Universe phenomena in principle, even if not feasible at present.
}

The remainder of this paper is structured as follows: In Section~\ref{section2}, we provide a brief overview of the evaporation formalism of PBHs. Section~\ref{section3} discusses the production of super-heavy sterile neutrinos from PBH evaporation in the early Universe and their subsequent decay into active neutrinos, generating an ultra-high-energy neutrino flux that can account for the KM3NeT observation. In Section~\ref{section4}, we explore the gravitational wave signatures of our framework, focusing on graviton production via Hawking radiation and graviton bremsstrahlung from sterile neutrino decay. {In Section~\ref{section5}, we discuss how the $N_{2,3}$ decay can generate matter-antimatter asymmetry in the early Universe through baryo-lepto-genesis route}, and finally conclude in Section~\ref{section6}.

\section{Primordial Black Hole Evaporation Formalism} \label{section2}
Primordial black holes form through gravitational collapse of early Universe overdensities, with initial mass $M_{\rm in}$ related to the background energy density $\rho_{\rm tot}$ and Hubble parameter $\mathcal{H}$ at formation temperature $T_{\rm in}$~\cite{Carr:1974nx,Carr_2010}:
\begin{equation}\label{eq:MassPBH}
    M_{\rm in} = \gamma \frac{4 \pi}{3} \frac{\rho_{\rm tot}(T_{\rm in})}{\mathcal{H}^3(T_{\rm in})},
\end{equation}
where $\gamma\simeq 0.2$, $\rho_{\rm tot} = 3 M_p^2 \mathcal{H}^2$, and $M_p \simeq 2.4 \times 10^{18}~\rm{GeV}$. The corresponding formation temperature $T_{\rm in}$ is related to $M_{\rm in}$ as: 
\begin{equation}\label{eq:initialTem}
    T_{\rm in} =\left(\frac{1440 \ \gamma^2}{g_*(T_{\rm in})}\right)^{1/4} M_p \sqrt{\frac{M_p}{M_{\rm in}}}\simeq 4.36 \times 10^{15}~\text{GeV} \left(\frac{1~\rm{g}}{M_{\rm in}}\right)^{1/2}.
\end{equation}
After formation, PBHs evaporate via Hawking radiation with temperature~\cite{Hawking:1974rv}:
\begin{equation}\label{eq:Hawkingtemperature}
    T_\text{BH} = \frac{M_p^2}{M_\text{BH}} \simeq 10^{13}~\rm GeV \left(\frac{1~\rm g}{M_{\rm BH}}\right).
\end{equation}
The energy spectrum for species $i$ with spin $s_i$ and mass $\mu_i$ produced via Hawking radiation is~\cite{UKWATTA201690}:
\begin{equation}\label{eq:energy_spectrum}
    \frac{d^2u_i(E,t)}{dE dt} = \frac{g_i}{2 \pi^2} \frac{\sigma_{s_i}(M_\text{BH}, \mu_i,E_i)}{e^{E_i/T_\text{BH}} -(-1)^{2s_i}} E_i^3,
\end{equation}
where $g_i$ is the number of degrees of freedom, $\sigma_{s_i}$ is the absorption cross-section. Thus, the mass-loss rate of PBH integrates over all species and can be written as:
\begin{equation}
    \frac{dM_\text{BH}}{dt} = - \varepsilon(M_\text{BH}) \frac{M^4_p}{M_\text{BH}^2},
\end{equation}
with evaporation function $\varepsilon(M_\text{BH}) \equiv \sum_i g_i \varepsilon_i (z_i)$ and:
\begin{equation}\label{eq:evap Func}
    \varepsilon_i (z_i) =\frac{27}{128\pi^3} \int_{z_i}^\infty \frac{\psi_{s_i}(x) (x^2 - z_i^2)}{e^x - (-1)^{2s_i}} x dx.
\end{equation}
In the geometric optics limit ($\psi_{s_i}=1$), the mass evolution follows:
\begin{equation}\label{eq:mass loss}
    \frac{d M_\text{BH}}{dt} \simeq - \frac{27\pi}{4} \frac{g_*(T_\text{BH})}{480} \frac{M_p^4}{M_\text{BH}^2}.
\end{equation}
Solving this equation gives PBH mass evolution:
\begin{equation}\label{eq:PBHmass}
    M_\text{BH}(t) = M_\text{in} \left(1- \frac{t-t_{i}}{\tau_{\rm BH}}\right)^{1/3},
\end{equation}
with the PBH lifetime given by:
\begin{equation}\label{eq:lifetime}
    \tau_{\rm BH}= \frac{4}{27} \frac{160 \ M_\text{in}^3}{\pi \ g_*(T_\text{BH}) M_p^4}  \simeq 2.66\times 10^{-28}~\rm{s}~\frac{100}{g_*(T_\text{BH})}\left(\frac{M_{\rm in}}{1~\rm{g}}\right)^{3}.
\end{equation}
The majority of PBH energy loss occurs in the final stages of evaporation. The evaporation temperature $T_\text{ev}$ depends on whether PBHs dominate the Universe pre-evaporation (matter-dominated, MD) or not (radiation-dominated, RD):
\begin{equation}\label{eq:evapTemperature}
  T_\text{ev} |_\text{MD} \simeq 3.55 \times 10^{10}~\text{GeV} \left(\frac{1~\text{g}}{M_\text{in}}\right)^{3/2},
\end{equation}
\begin{equation}\label{eq:ev temperaturePBH}
    T_\text{ev} |_\text{RD} \simeq \frac{\sqrt{3}}{2} T_\text{ev} |_\text{MD}.
\end{equation}
The allowed range of PBH masses is constrained by the maximum Hubble scale after inflation ($\mathcal{H}< 2.5 \times 10^{-5} M_p$) and the requirement that PBH evaporation does not disrupt Big Bang Nucleosynthesis (BBN), necessitating $T_{\rm ev} > T_{\rm BBN} \simeq 4$ MeV~\cite{Planck:2018jri}. These constraints yield~\cite{Carr:2020gox}:  
\begin{equation}
0.4 \ \text{g} \lesssim M_\text{in} \lesssim 9.7 \times 10^8 \ \text{g}.
\end{equation}

\section{Early Universe Sterile Neutrino Decay as the Source of KM3-230213A}\label{section3}
To explain the observed ultra-high-energy neutrino event detected by KM3NeT, we propose a novel cosmological production mechanism involving PBHs in the early Universe. While light PBHs (with masses $M_{\rm PBH}\lesssim 10^9$g) evaporate completely before neutrino decoupling, they can serve as efficient source of production for super-heavy long-lived particles through Hawking radiation. These particles, which we identify as sterile neutrinos in our model, are produced independently of their coupling strength to the Standard Model sector due to the gravitational nature of PBH evaporation.

The key insight of our scenario is this two-stage process:
\begin{enumerate}
    \item PBHs produce superheavy sterile neutrinos ($N_1$) via Hawking evaporation.
    \item These $N_1$ particles subsequently decay into active neutrinos with energies $E_\nu \sim M_{N_1}/2$
\end{enumerate}

The resulting neutrino flux experiences significant redshift as the Universe expands, with a fraction arriving at present-day Earth with energies $\mathcal{O}(100)$ PeV precisely matching the KM3-230213A observation. This cosmological origin naturally explains the extragalactic, isotropic nature of the ultra-high-energy neutrino flux, aligning with the fact that the detected signal originates from a direction nearly opposite to the Galactic Center. Thus, it offers a compelling alternative to conventional astrophysical explanations, with distinctive testable predictions for both multi-messenger observations and future high-energy neutrino experiments. 

\subsection{Neutrino Mass and Sterile Neutrino Couplings}\label{sec:nucouplings}
{The relevant Lagrangian for neutrino mass is given by
\begin{eqnarray}
    \mathcal{L} \supset -Y_{\alpha k}\overline{L}_\alpha\tilde{H} N_k- \frac{1}{2} (\bar{N^c_k}M_{N_k} N_k)+ {\rm h.c.}
\end{eqnarray}
The neutrino mass is given as,
\begin{eqnarray}
    \left(m_\nu\right)_{\alpha \beta}=-M_DM_R^{-1}M_D^T\equiv -\frac{1}{2}Y_{\alpha k} M_{N_k}^{-1} Y_{k \beta} v^2
    \label{as1}
\end{eqnarray}
where $v= \langle H \rangle$ is the vacuum expectation value of the SM Higgs. 
We use the Casas-Ibarra parametrization~\cite{Casas:2001sr} to calculate the Yukawa couplings that satisfy the neutrino oscillation data as:
\begin{eqnarray}
    Y=\frac{i \sqrt{2}}{v}~ U^* ~D_{\sqrt{m_\nu}}~R^T~D_{\sqrt{M_R}}
\end{eqnarray}
where $U$ is the PMNS matrix, $R$ is an arbitrary complex orthogonal matrix satisfying $R^T R=I$ that can be parametrized in terms of three complex angles $z_{ij}$, and $D_{\sqrt{m_\nu}}$ and $D_{\sqrt{M_R}}$ are the diagonal matrices for the square roots of the light neutrino masses and RHN masses, respectively. Here, it is worth mentioning that, with an appropriate choice of the $R$ matrix, the Yukawa couplings of $N_1$ can be made dependent on the lightest active neutrino mass whereas $N_2$ and $N_3$ couplings are dominantly used to explain the oscillation data. By choosing a vanishingly small value for the lightest active neutrino mass, it is possible to obtain the desirable small couplings of $N_1$, ensuring that it decays after neutrino decoupling and produces the ultra-high-energy neutrino flux needed to explain the KM3NeT event.
}


Another crucial consideration for models involving fermionic extensions of the SM with very heavy degrees of freedom and large couplings is their potential impact on the stability of the SM electroweak vacuum. The introduction of RHNs can destabilize the vacuum by contributing negatively to the Renormalization Group (RG) evolution of the Higgs quartic coupling, $\lambda$~\cite{Fu:2023nrn, Ipek:2018sai}. Furthermore, the RHN masses must remain compatible with the perturbativity of the Yukawa couplings while explaining the observed light neutrino masses~\cite{Ipek:2018sai}. For RHNs with masses approaching the Planck scale, as considered here, a complete analysis must also account for gravitational corrections to the vacuum decay rate~\cite{Chauhan:2023pur}.
In our framework, this analysis is straightforward. The heaviest RHN, $N_1$, despite its near-Planck scale mass, possesses extremely suppressed Yukawa couplings ($|Y_{\alpha 1}| \sim 10^{-21}$) and is therefore physically decoupled from the RG evolution~\cite{Ipek:2018sai}. The contributions to the beta functions, which depend on terms like $\text{Tr}(Y^\dagger_\nu Y_\nu)$ and $\text{Tr}(Y^\dagger_\nu Y_\nu Y^\dagger_\nu Y_\nu)$, are negligible for $N_1$. Consequently, the question of vacuum stability is entirely determined by the two lighter states, $N_2$ and $N_3$, which have sizable Yukawa couplings to generate the observed light neutrino masses.

For the purpose of stability, our model is thus effectively a two-RHN high-scale seesaw scenario.  Even in the most conservative case considered in~\cite{Chauhan:2023pur}, which neglects the stabilizing effects of gravity (corresponding to a non-minimal Higgs-gravity coupling $\xi = -1/6$), the upper bounds on the RHN masses for a Normal Hierarchy of light neutrinos are $M_{\text{heavier}} \lesssim 10^{15.3}$~GeV and $M_{\text{lighter}} \lesssim 10^{14.4}$~GeV~\cite{Chauhan:2023pur}. In our analysis, we ensure that the chosen masses for $N_3$ and $N_2$ lie comfortably within these allowed limits~\cite{Ipek:2018sai, Chauhan:2023pur}. Furthermore, as demonstrated in Ref.~\cite{Chauhan:2023pur}, including gravitational corrections (for $\xi \geq 0$) acts to stabilize the vacuum, further relaxing these bounds. We therefore conclude that our proposed scenario is robustly consistent with the requirement of a metastable electroweak vacuum.

To explicitly demonstrate the viability of our setup, let us consider a diagonal RHN mass matrix,
\begin{equation}\label{eq:rhnMR}
M_R = \text{Diag}\{m_{N_1}, m_{N_2}, m_{N_3}\} = \text{Diag}\{0.1 M_p,\, 2.5 \times 10^{-5} M_p,\, 2.5 \times 10^{-4} M_p\}.
\end{equation}
The orthogonal matrix $R$ is parametrized as a rotation in the 2-3 plane with a complex mixing angle $z_{23}= \theta_{R} + i \theta_I$. For simplicity, we assume the two Majorana phases in $U$ to be zero as the  Majorana phases are currently
completely unconstrained. We adopt the best-fit values of the neutrino oscillation parameters~\cite{ParticleDataGroup:2024cfk,Esteban:2020cvm}, with $\theta_{R,I} \in [-\pi,\pi]$. Using the Casas-Ibarra parametrization, the Yukawa coupling matrix for a benchmark choice $z_{23}= -0.9426 \pi + i\, 0.06978 \pi$ is then obtained to be:

\begin{eqnarray}\label{eq:yukawa}
    Y= \begin{pmatrix}
        0. + 2.32488\times10^{-21} i&0.0100373 - 0.0604792 i&-0.0003079 - 0.18223 i\\
        7.7151\times10^{-23} - 7.62991\times10^{-22} i&0.0504647 + 0.123094 i&0.085298 - 0.706361 i\\
        6.68303\times10^{-23} + 1.39073\times10^{-21} i&-0.0494323 + 0.0387449 i&0.0253041 + 0.695677 i\\
    \end{pmatrix} \nonumber \\
\end{eqnarray} 
As can be seen, it is possible to have the $N_1$ couplings of $\mathcal{O}(10^{-21})$ while $N_2$ and $N_3$ have large couplings of order $\mathcal{O}({10^2})$ or larger. Considering the decay widths for the dominant tree level decay channels of $N_1$~\cite{Higaki:2014dwa}, the sterile neutrino lifetime is given by:  
\begin{equation}\label{eq:SN_lifetime}
    \tau_{N_1} \simeq  68 \ {\rm s} \left(\frac{10^{-21}}{|Y_{\alpha 1}|}\right)^2 \left(\frac{0.1 M_p}{m_{N_1}}\right),
\end{equation}
where $|Y_{\alpha 1}|$ is the Yukawa coupling associated with the interaction $|Y_{\alpha 1}| \overline{L}_\alpha \tilde{H} N_1$.
It should be noted that for $m_{N_1} \gg  m_h,m_W,m_Z$, the decay widths have the ratio:  $ \Gamma_{N_1 \rightarrow \nu_\alpha h}:\Gamma_{N_1 \rightarrow \nu_\alpha Z}:\Gamma_{N_1 \rightarrow \ell^{-}_\alpha W^+}=1:1:2$. This ratio is relevant for estimating the neutrino flux from $N_1$ decay, which we will explore in the next section.

\subsection{Heavy Sterile Neutrino Production from PBH}
PBHs can emit all particles in the spectrum through Hawking evaporation, including gravitons and heavy sterile neutrinos, provided their masses satisfy $m_i \lesssim T_{\rm BH}$. For a Schwarzschild black hole, the greybody factor in the geometric optics limit is approximated as $\sigma_{s_i} = (27/64\pi) M_{\rm BH}^2/M_p^4$. Consequently, the emission in Eq.~(\ref{eq:energy_spectrum}) spectrum simplifies to~\cite{UKWATTA201690,Lunardini:2019zob,Perez-Gonzalez:2020vnz}:  
\begin{equation}
    \frac{d^2 u_{i}}{dtdE} \simeq \frac{27 g_i}{64 \pi^3} \frac{M_{\rm BH}^2}{M_p^4} \frac{E_i^3}{e^{E_i/T_{\rm BH}}\pm 1},
\end{equation}
where the $(+)$ and $(-)$ signs correspond to fermion and boson production, respectively.

Defining the comoving energy densities of radiation and PBH as $\tilde{\rho}_r = a^4 \rho_r$ and $\tilde{\rho}_{\rm BH} = a^3 \rho_{\rm BH}$, the Boltzmann equations governing their evolution are~\cite{Masina:2020xhk,Giudice:2000ex,Bernal:2020bjf,JyotiDas:2021shi,Barman:2021ost}:  
\begin{equation}
\begin{aligned}
    & \frac{dM_\text{BH}}{d \ln(a)} = - \frac{\varepsilon(M_\text{BH})}{\mathcal{H}} \frac{ M_p^4}{M_\text{BH}^2}, \\
    & \frac{d \tilde{\rho}_\text{BH}}{d \ln(a)} =  \frac{\tilde{\rho}_\text{BH}}{M_\text{BH}} \frac{dM_\text{BH}}{d \ln(a)}, \\
    & \frac{d \tilde{\rho}_r}{d \ln(a)} = -\frac{\varepsilon_\text{SM}(M_\text{BH})}{\varepsilon(M_\text{BH})} \frac{a \ \tilde{\rho}_\text{BH}}{M_\text{BH}} \frac{dM_\text{BH}}{d \ln(a)},
\end{aligned}
\end{equation}\label{Bolteq}
{where $\mathcal{H}$ is the Hubble expansion rate expressed as $\mathcal{H}={\dot a}/{a}=(\sqrt{3}M_p)^{-1} \rho_{\rm tot}$.}
Here, $\varepsilon_\text{SM}\equiv g_{\rm SM}\sum_i \varepsilon_i (z_i)$ accounts for the evaporation contributions from Standard Model particles, excluding gravitons.  As we assume a monochromatic PBH mass spectrum, the comoving PBH number density remains conserved:  
\begin{equation}\label{eq:nbhconst}
    n_{\rm BH} (t) = n_{\text{BH}}(t_i) \left(\frac{a_i}{a}\right)^3,
\end{equation}
where the initial number density $n_{\rm BH}(t_i)$ is given by:  
\begin{equation}\label{eq:nBHin}
    n_{{\rm BH}} (t_i) = \beta  \frac{\rho_{\rm r} (t_i)}{M_{\rm in}} = \beta \frac{48 \pi^2 \gamma^2 M_p^6}{M_{\rm in}^3}.
\end{equation}
Here, $\beta$ is the initial energy density of PBHs relative to radiation {\it i.e.} $\beta={\rho_{\rm BH}(T_{\rm in})}/{\rho_{r}(T_{\rm in})}$.
Once produced via PBH evaporation, the heavy sterile neutrino $N_1$ can decay into ultra-high-energy neutrinos and other SM particles. The Boltzmann equation tracking its evolution is:  
\begin{equation}\label{eq:boltzeqN} 
     \frac{d \tilde{n}_{N_1}}{d \ln(a)} =  \frac{\tilde{\rho}_\text{BH}}{M_\text{BH}} \frac{\Gamma_{\text{BH}\rightarrow N_1}}{\mathcal{H}} - \frac{{\Gamma_{N_1}}}{\mathcal{H}}  \tilde{n}_{N_1},
\end{equation}
which has to be solved along with Eq.~(\ref{Bolteq}) simultaneously.
Here $\tilde{n}_{N_1} \equiv n_{N_1} a^3$ is the comoving number density of sterile neutrinos, and {$\Gamma_{N_1}$ is the total decay rate of $N_1$. 
The momentum-integrated sterile neutrino emission rate from PBH evaporation is:  
\begin{equation}
        \Gamma_{\text{BH}\rightarrow N_1}(t) = \frac{27 g_{N_1}}{128 \pi^3} \frac{M_p^2}{M_\text{BH}(t)}\int_{z}^\infty \frac{\psi_{N_1}(x) (x^2 - z^2)}{e^x - 1}dx,
\end{equation}
where $x=E/T_\text{BH}$ and $z = m_{N_1}/T_{\rm BH}$. In the geometric optics limit, the emission rate simplifies to~\cite{Cheek2022,Perez-Gonzalez:2020vnz}:  
\begin{equation} \label{eq:PBHdecay}
\Gamma_{\text{BH}\rightarrow N_1}(t) = \frac{27 g_{N_1}}{64 \pi^3} \frac{M_p^2}{M_\text{in}} \left(1-\frac{t-t_i}{\tau}\right)^{-1/3}\mathcal{F}(z),
\end{equation}
where $\mathcal{F}(z) = [z  {\rm Li}_2(e^{-z}) + {\rm Li}_3(e^{-z})]$, with ${\rm Li}_n$ denoting the polylog function of order $n$. In the high-temperature limit ($z\ll1$), $\mathcal{F}(z) \to \zeta(3) \approx 1.20206$. For our numerical analysis, we utilize the "ULYSSES" package~\cite{Granelli:2020pim}, incorporating modified graybody factors{\footnote{In this paper, we adhere to the semi-classical description of black hole evaporation, as implemented in the numerical package ULYSSES. We do not consider quantum effects that suggest black hole evolution may be altered after a certain time due to phenomena such as the memory burden effect~\cite{Dvali:2020wft,Dvali:2024hsb} and Page time~\cite{Page:2013dx,Perez-Gonzalez:2025try}, as a complete determination of these effects is yet to be achieved.}. 

\begin{figure}[htbp]
    \centering
    \includegraphics[width=0.495\textwidth]{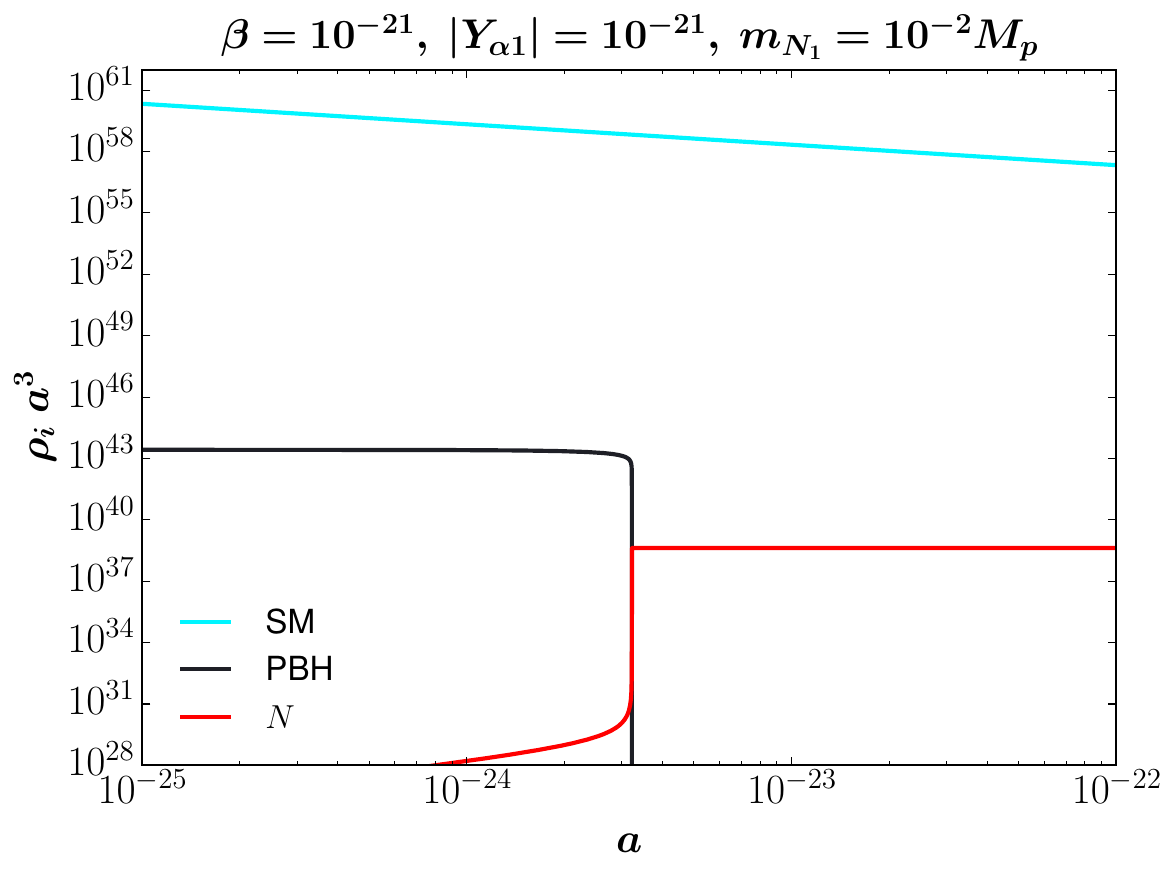}
     \includegraphics[width=0.495\textwidth]{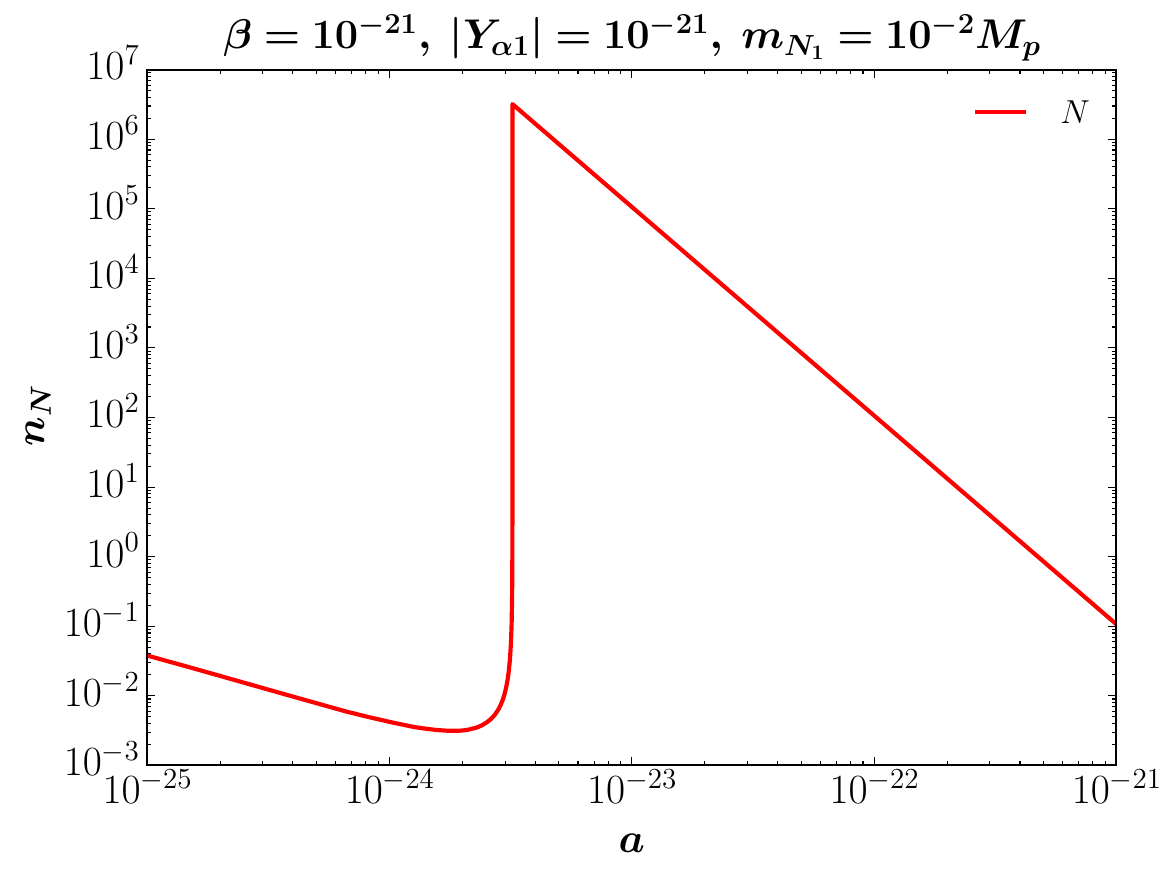}
     \includegraphics[width=0.495\textwidth]{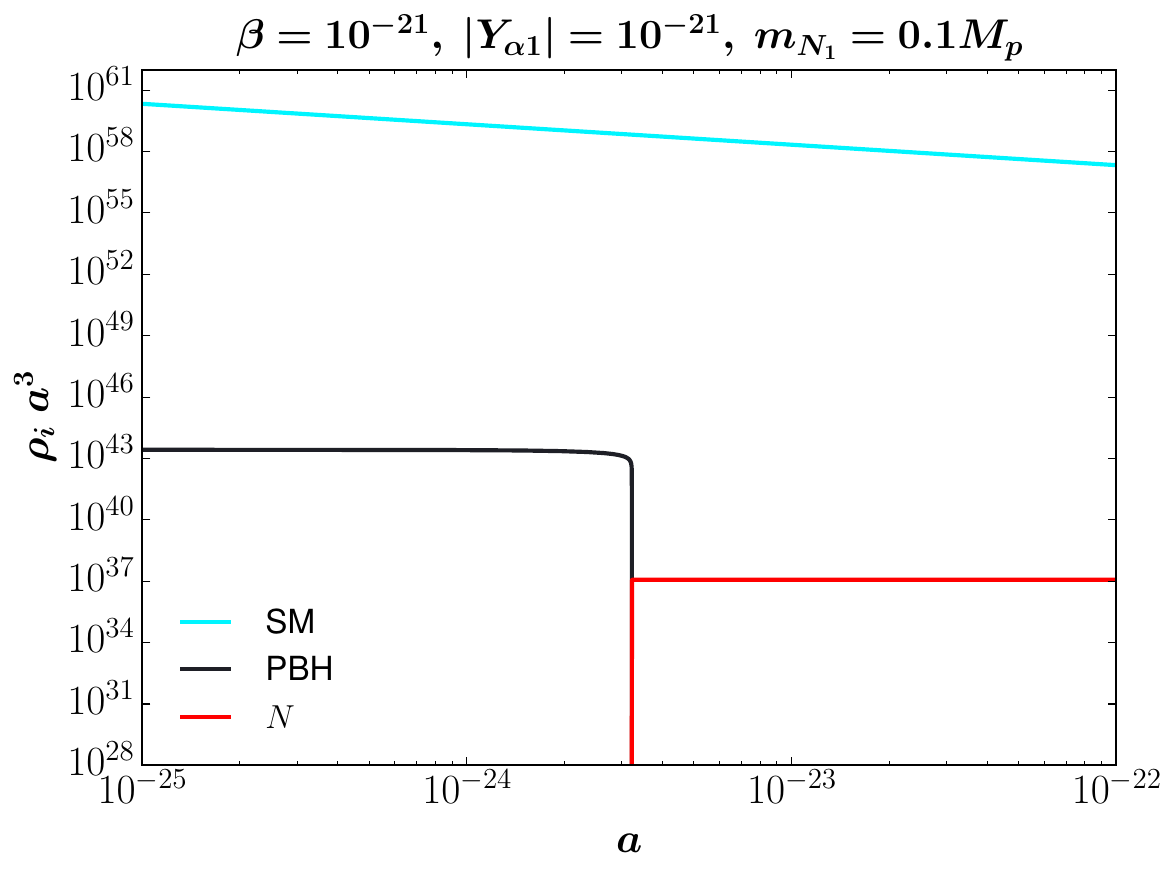}
     \includegraphics[width=0.495\textwidth]{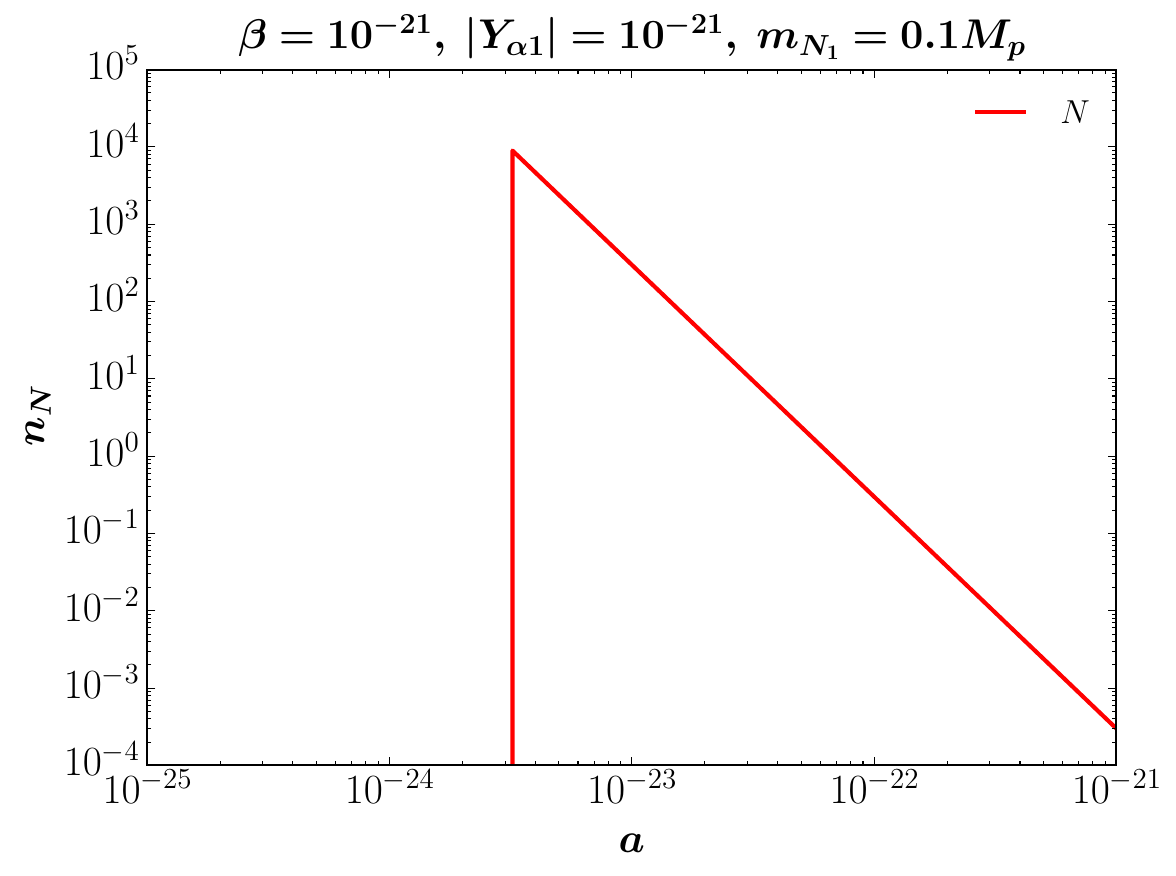}
    \caption{Evolution of the comoving energy densities for radiation, PBH, and sterile neutrinos (left) and the sterile neutrino number density (right) as functions of the scale factor for two different sterile neutrino masses, $m_{N_1}=10^{-2} M_p$ and $0.1 M_p$. Here, we take the initial PBH mass as $M_{\rm in} = 1$ g.}
    \label{fig:evolution}
\end{figure}

Fig.~\ref{fig:evolution} illustrates the numerical evolution of the comoving energy densities for radiation (cyan), PBHs (black), and sterile neutrinos (red) in the left panel, while the right panel illustrates the evolution of the sterile neutrino number density. Due to the chosen value of $\beta$, PBHs never dominate the Universe, ensuring that both their formation and evaporation occur during the radiation-dominated era. 

As PBHs evaporate, their temperatures increase, potentially reaching the Planck scale. Consequently, at the final stage of evaporation, PBHs can emit extremely heavy particles. If the mass of such a particle is large, its emission remains suppressed until $T_{\rm BH}$ surpasses the particle mass. This effect is clearly visible in the evolution of $N_1$ energy density and number density, where a sharp enhancement appears around $a \sim 10^{-24}$, signaling that the majority of sterile neutrinos are produced near the end of PBH evaporation when $T_{\rm BH} \gtrsim m_{N_1}$. We define the time at which $T_{\rm BH}=m_{N_1}$ as $t_1$, corresponding to a PBH mass of $M_{\rm BH} (t_1) = {M_p^2}/{m_{N_1}}$, allowing us to express $t_1$ as:
\dis{
t_1 = t_i+ \tau \left[1-\bfrac{M_p^2}{M_{\rm in}m_{N_1}}^3\right].
}

In this work, we focus on a scenario where the sterile neutrino is long-lived, decaying well after neutrino decoupling. The Yukawa coupling is constrained by requiring $\tau_{N_1}>\tau_{\rm \nu-{\rm decoupling}}\simeq 1 \ \rm{s}$:
\dis{
|Y_{\alpha 1}| < 8.2 \times 10^{-21} \bfrac{0.1 M_p}{m_{N_1}}^{1/2}.
}
In this regime, the second term on the right-hand side of Eq.~(\ref{eq:boltzeqN}) can be neglected when considering sterile neutrino production from PBHs. The sterile neutrino number density $n_{N_1}$, immediately following PBH evaporation, is given by:
\begin{equation}
    (a^3 n_{N_1})|_{\rm ev} = n_{\rm BH}(t_i) a_i^3 \int_{t_i}^{t_{\rm ev}}\Gamma_{BH\rightarrow N_1}(t)\,dt,
\end{equation}
where we use the fact that the comoving number density of PBHs remains conserved, i.e., $n_{\rm BH}(t_i) a_i^3 = n_{\rm BH}(t_{\rm ev}) a_{\rm ev}^3$. Therefore, it can be estimated as:
\begin{eqnarray}\label{eq:nchiatev}
    n_{N_1}(a_{\rm ev}) &=& \frac{27 g_{N_1}}{64 \pi^3}  \frac{M_p^2}{M_{\rm{in}}} n_{\rm BH}(t_i) \bfrac{a_i}{a_{\rm ev}}^3  \int_{t_1}^{t_{\rm ev}} dt \left(1-\frac{t-t_i}{\tau}\right)^{-1/3}~\mathcal{F}(z),
     \nonumber \\
    &\simeq& \frac{15~g_{N_1} ~\zeta(3)}{\pi^4g_*(T_{\rm BH}) } \frac{M^2_p}{m^2_{N_1}} n_{\rm BH}(t_i) \bfrac{a_i}{a_{\rm ev}}^3 \simeq 0.2~ \beta \frac{M_p^{11}}{m_{N_1}^2 M_{\rm in}^6}   \nonumber \\
    &\simeq& { 4 \times 10^{3} \ \text{GeV}^{3} \left(\frac{\beta}{10^{-21}}\right) \left(\frac{1 \ \rm{g}}{M_{\rm in}}\right)^6\left({\frac{0.1 M_{p}}{m_{N_1}}}\right)^2},
\end{eqnarray}
where Eq.~(\ref{eq:nBHin}) is used for $n_{\rm BH}(t_i)$, and $\mathcal{F}(z)\simeq \zeta(3)$ is approximated for $T_{\rm BH} > m_{N_1}$ in the second line. The ratio of scale factors $a_i/a_{\rm ev}$ is determined using entropy conservation, along with Eqs.~(\ref{eq:initialTem}) and (\ref{eq:evapTemperature}), and is given by:
\dis{
\frac{a_i}{a_{\rm ev}} = \frac{T_{\rm ev}}{T_{\rm in}} \left(\frac{g_{*,s}(T_{\rm ev})}{g_{*,s}(T_{\rm in})}\right)^{1/3} \simeq 1.5 \frac{M_p}{M_{\rm in}}.
}

\subsection{Neutrino Flux from Sterile Neutrino Decay}

In this work, we analyze the production of high-energy neutrinos via delayed decays of sterile neutrinos ($N_1$) which are produced in primordial black hole evaporation. The sterile neutrinos, produced non-thermally via Hawking radiation, decay into active neutrinos ($N_1\to h\nu$) after neutrino decoupling. The resulting neutrino flux is shaped by cosmological redshift and the decay kinematics, which can be quantified through the Boltzmann equation~\cite{Wu:2024uxa}: 
\dis{
\left[\frac{\partial}{\partial t} - \mathcal{H}~ p ~\partial_p\right] f(t, p) = (1-f) \Gamma_{\rm prod} - f \Gamma_{\rm abs},
}
with $\Gamma_{\rm prod}$ and $\Gamma_{\rm abs}$ the production and absorption rates of the species. In the limit of $f\ll 1$, the analytic solution of the distribution function can be found as: 
\dis{\label{eq:distfunc}
f(t,p) = \int_0^a \frac{\Gamma_{\rm prod}(a', p')}{\mathcal{H}(a') a'} da', \quad {\rm where}~~~~p' \equiv p\frac{a}{a'}\,.
}
Considering, {$N_1\rightarrow  h(Z)+\nu$ ($1\rightarrow 2+3$) with $m_{N_1}\gg m_\nu, m_{h(Z)}$}, the production rate becomes:
\begin{eqnarray}
    \Gamma_{\rm prod}^{(f_3)} &=& \frac{1}{2 E_3} \int \frac{d^3 \boldsymbol{p}_1}{(2 \pi)^3 2 E_1} \frac{d^3 \boldsymbol{p}_2}{(2 \pi)^3 2 E_2} f_1 (2 \pi)^4 \delta^4(p_1^\mu-p_2^\mu-p_3^\mu) |\mathcal{M}|^2 \nonumber \\
    &=& \frac{1}{2 E_3} \int \frac{ 2\pi p_1^2 d {p}_1 dc_{13}}{(2 \pi)^3 4 E_1 E_2} f_1 |\mathcal{M}|^2 (2 \pi) \delta(E_1 - E_2 -E_3)|_{E_2\rightarrow\sqrt{p_1^2 + p_3^2 - 2 p_1 p_3 c_{13}}}, 
\end{eqnarray}
where $c_{13} \equiv (\boldsymbol{p}_1 \cdot \boldsymbol{p}_3)/(p_1 p_3)$ {with $p_1=|\boldsymbol{p}_1|,  p_3=|\boldsymbol{p}_3|$ } and $|\mathcal{M}|^2$ is the squared amplitude of the decay process. For the case where sterile neutrino follows a non-relativistic distribution and the squared matrix element $|\mathcal{M}|^2$ can be treated as constant (and thus factored out of the integral), we can simplify the analysis by noting that $E_1 = m_1$ and $p_1=0$. Consequently, the energy of the daughter neutrino becomes $E_2\rightarrow p_\nu=E_\nu$.
After performing the integration over the angular variable 
$c_{13}$ and accounting for the energy-conserving delta function, the production rate $\Gamma_{\rm prod}^{(f_3)}$ simplifies to: 
\begin{eqnarray}
    \Gamma_{\rm prod}^{(f_3)} 
    &\simeq& \frac{\pi |\mathcal{M}|^2}{2 m_1^3} n_1 \delta(E_3 - \frac{m_1}{2}),
\end{eqnarray}
where $n_1 = \int f_1 \frac{d^3 \boldsymbol{p}_1}{(2 \pi)^3}$ is the number density of particle 1 {\it i.e.} the sterile neutrino. 
Since the sterile neutrino decay occurs in a non-relativistic regime, the production rate $\Gamma_{\rm prod}$ for neutrinos is 
a delta function, and thus Eq.~(\ref{eq:distfunc}) can be written as:
\begin{eqnarray}\label{eq:fnu}
    f_\nu &\simeq& \frac{\pi |\mathcal{M}|^2}{2 m_{N_1}^3} \int_0^a \frac{n_{N_1}(a')}{\mathcal{H}(a') a'} \delta(\frac{E_\nu a}{a'} - \frac{m_{N_1}}{2}) da' \simeq \frac{\pi |\mathcal{M}|^2}{m_{N_1}^4} \frac{n_{N_1}(\tilde{a})}{\mathcal{H}(\tilde{a})},
\end{eqnarray}
with $\tilde{a} = 2 E_\nu a/m_{N_1}$. Assuming that the sterile neutrinos decay with a rate $\Gamma_{N_1}$ in a radiation-dominated Universe, the evolution of the Hubble parameter $\mathcal{H}(a)$ and the sterile neutrino number density $n_{N_1}(a)$ follows:
\dis{ \label{eq:funcof_a}
\mathcal{H}(a) = \mathcal{H}_{*} \left(\frac{a_{*}}{a}\right)^2, \quad n_{N_1}(a) = n_{N_1*} \left(\frac{a_{*}}{a}\right)^3 \exp\left[-\frac{\Gamma_{N_1}}{2} \left(\frac{1}{\mathcal{H}(a)}-\frac{1}{\mathcal{H}_{\rm *}}\right) \right],
}
where $a_*$ corresponds to the epoch when the sterile neutrino starts to decay. Substituting Eq.~(\ref{eq:funcof_a}) into Eq.~(\ref{eq:fnu}), we obtain 
\begin{eqnarray}\label{eq:fnu_wrt_r}
    f_\nu &\simeq& \frac{\pi |\mathcal{M}|^2}{m_{N_1}^4} \frac{n_{N_1*}}{\mathcal{H}_{*}} \left(\frac{2 E_\nu a/m_{N_1}}{a_{*}}\right)^2 \left(\frac{a_{*}}{2 E_\nu a/m_{N_1}}\right)^3 \exp\left[- \frac{\Gamma_{N_1}}{2} \left(\frac{(2E_\nu a/m_{N_1})^{2}}{\mathcal{H}_{*} a_{*}^2} - \frac{1}{\mathcal{H}_{*}}\right)\right] \nonumber \\
    &\simeq& \frac{\pi |\mathcal{M}|^2 n_{N_1*} a_{*}^3}{m_{N_1} \mathcal{H}_{*}} \left(\frac{2 E_\nu a}{m_{N_1} a_{*}}\right)^2 \left(\frac{1}{2 E_\nu a}\right)^3 \exp\left\{-\frac{1}{2} \frac{\Gamma_{N_1}}{\mathcal{H}_{*}} \left[\left(\frac{2 E_\nu a}{m_{N_1} a_{*}}\right)^{2}-1\right]\right\}.
\end{eqnarray}
Considering the total decay rate formalism for the two-body decay process, $\Gamma_{N_1} = {|\mathcal{M}|^2}/{16 \pi m_{N_1}}$, Eq.~(\ref{eq:fnu_wrt_r}) is simplified as follows:
\dis{
f_\nu \simeq 16 \pi^2 \frac{\Gamma_{N_1}}{\mathcal{H}_{*}} n_{N_1*} a_{*}^3 \left(\frac{2 E_\nu a}{m_{N_1} a_{*}}\right)^2 \left(\frac{1}{2 E_\nu a}\right)^3 \exp\left\{-\frac{1}{2} \frac{\Gamma_{N_1}}{\mathcal{H}_{*}} \left[\left(\frac{2 E_\nu a}{m_{N_1} a_{*}}\right)^{2}-1\right]\right\}.
}
As the sterile neutrino number density remains conserved until its decay at scale factor $a_{N_1}$, where $\mathcal{H}* = \Gamma_{N_1}$, it follows that $n_{N_1*} a_*^3 = n_{N_1}(a_{\rm ev}) a_{\rm ev}^3$. Utilizing this, the above equation is written by:
\begin{equation}
    f_\nu(t, E_\nu) \simeq 16 \pi^2 n_{N_1}(a_{\rm ev}) a_{\rm ev}^3  \left(\frac{2 E_\nu a(t)}{m_{N_1} a_{N_1}}\right)^2 \left(\frac{1}{2 E_\nu a(t)}\right)^3 \exp\left\{-\frac{1}{2} \left[\left(\frac{2 E_\nu a(t)}{m_{N_1} a_{N_1}}\right)^{2}-1\right]\right\}.\\
\end{equation}
Thus redshifting to the present day,  we obtain the distribution to be:  
\begin{equation}\label{eq:fnutoday}
   f_\nu(E_\nu)|_{t\rightarrow t_0} 
   \simeq 16 \pi^2 n_{N_1}(a_{\rm ev}) a_{\rm ev}^3  \left(\frac{2 E_\nu}{m_{N_1}} \frac{a_0}{a_{N_1}}\right)^2 \left(\frac{1}{2 E_\nu a_{0}} \right)^3 \exp\left\{-\frac{1}{2} \left[\left(\frac{2 E_\nu a_{0}}{m_{N_1} a_{N_1}}\right)^{2}-1\right]\right\}.\,
\end{equation}
The temperature at which the sterile neutrino decays denoted as $T_{N_1}$, can be determined during the radiation-dominated era using the condition $H=1/2 \tau_{N_1}$ along with Eq.~(\ref{eq:SN_lifetime}). This yields:
\begin{eqnarray}\label{eq:TemSN}
    T_{N_1} \simeq \frac{3 |Y_{\alpha 1}|}{4 \pi} \left(\frac{M_p m_{N_1}}{g_*^{1/2}(T_{N_1})}\right)^{1/2} = 0.13 \ {\rm{MeV}} \left(\frac{|Y_{\alpha 1}|}{10^{-21}}\right) \left(\frac{4}{g_*(T_{N_1})}\right)^{1/4} \left(\frac{m_{N_1}}{0.1  M_p}\right)^{1/2}
\end{eqnarray}
  Consequently, the ratio of the scale factor at the time of decay to its present value, $a_{N_1}/a_0$, is estimated as:  
\begin{eqnarray}\label{eq:axtoa0}
    \frac{a_{N_1}}{a_0} \simeq \frac{T_0}{T_{N_1}} \left(\frac{g_{*,s}(T_0)}{g_{*,s}(T_{N_1})}\right)^{1/3} \simeq 1.7 \times 10^{-9}  \left(\frac{10^{-21}}{|Y_{\alpha 1}|}\right) \left(\frac{0.1 M_p}{m_{N_1}}\right)^{1/2}
\end{eqnarray}
where we have used $ g_{*,s}(T_{N_1}) \simeq g_{*}(T_{N_1}) \simeq 4$ in the last step.

Using the result of Eqs.~(\ref{eq:nchiatev}), and (\ref{eq:axtoa0}) in Eq.~(\ref{eq:fnutoday}), we obtain the the present day total differential flux of neutrinos per unit solid angle, for a single flavour to be:
\begin{eqnarray}
     \frac{d^2\Phi_\nu}{dE_\nu d \Omega} &=& \frac{{\rm BR}_{N_1\to\nu}}{3}\frac{1}{4 \pi}\frac{E_\nu^2}{2 \pi^2} f_\nu(E_\nu)|_{t\rightarrow t_0}=\frac{1}{2\times3}{\frac{1}{4 \pi}}\frac{E_\nu^2}{2 \pi^2} f_\nu(E_\nu)|_{t\rightarrow t_0} \nonumber \\
    &=& \frac{E_\nu}{{6\pi} m_{N_1}^2} n_{N_1}(a_{\rm ev}) \bfrac{a_{\rm ev}}{a_0}^3 \left(\frac{a_{N_1}}{a_0}\right)^{-2} \exp\left\{- \frac{1}{2} \left[\left(\frac{2 E_\nu}{m_{N_1}}\right)^{2}\left(\frac{a_{N_1}}{a_0}\right)^{-2} -1\right]\right\} \nonumber \\
    &\simeq& 
    {5.5} \times 10^{-27} ~\text{GeV}^{-1} \ \text{cm}^{-2} \ \text{s}^{-1} \ \text{sr}^{-1}\left(\frac{E_\nu}{10^{8}\ \rm{GeV}}\right) \left(\frac{\beta}{10^{-21}}\right) \left(\frac{|Y_{\alpha 1}|}{10^{-21}}\right)^{2} \left(\frac{0.1 M_p}{m_{N_1}}\right)^3  \nonumber \\
    &\times& \left(\frac{M_{\rm in}}{1 \ \rm{g}}\right)^{1/2} \exp\left\{- \frac{1}{2} \left[ 0.22 \left(\frac{E_\nu}{10^{8} \ \rm{GeV}}\right)^2 \left(\frac{|Y_{\alpha 1}|}{10^{-21}}\right)^{2} \left( \frac{0.1 M_p}{m_{N_1}}\right)-1\right] \right\}.
\end{eqnarray}

\begin{figure}[tbp]
    \centering
     \includegraphics[width=\textwidth]{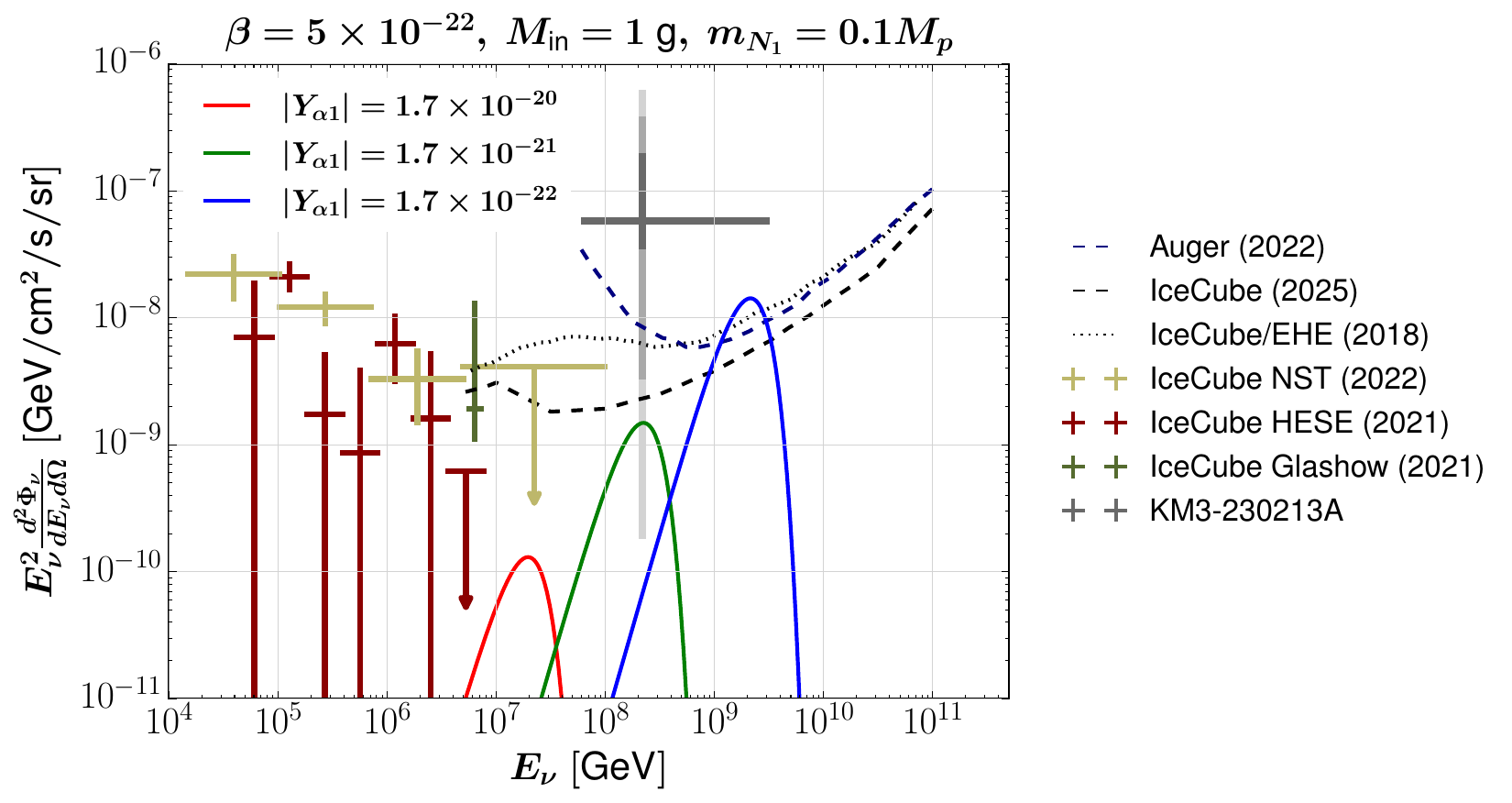}
    \caption{Predicted single flavor neutrino flux from decays of superheavy sterile neutrinos ($m_{N_1}=0.1 M_p$), produced via PBH evaporation, shown for PBH mass $M_{\rm in}=1 \ \rm{g}$ and $\beta=5 \times 10^{-22}$. Different colored curves represent different Yukawa couplings governing the sterile neutrino decay rate. 
   }
    \label{fig:neutrinoflux}
\end{figure}

Fig.~\ref{fig:neutrinoflux} illustrates the predicted single-flavor neutrino flux arising from the decay of heavy sterile neutrinos emitted during the evaporation of PBHs. The neutrino flavour ratio at Earth is assumed
to be $1:1:1$. {Since we are specifically calculating the muon neutrino flux to explain the observed muon event at KM3NeT, a factor of $1/3$ appears in the flux expression.} {As mentioned in the previous section, neutrinos are primarily produced through the decay channels $N_1\to \nu h$ and $N_1\to \nu Z$, so we account for the combined branching ratios of these modes when calculating the flux {leading to an additional factor of $1/2$ in the expression.}} The flux is shown for three representative values of the Yukawa coupling, while keeping the initial PBH mass $M_{\rm in}=1g$, $\beta=5 \times 10^{-22}$, and the sterile neutrino mass $m_{N_1}=0.1 M_p$ fixed. The shape and normalization of the flux crucially depend on the chosen Yukawa coupling, which controls the lifetime of the sterile neutrino and thus the redshifted energy distribution of the resulting neutrinos. Notably, the flux peaks in the ultra-high-energy regime, with a characteristic spectral cutoff determined by the sterile neutrino decay kinematics.

For comparison, relevant experimental data and limits are shown: KM3-230213A event~\cite{KM3NeT:2025npi} (gray points), IceCube single power-law fits from the 7.5-year HESE dataset~\cite{IceCube:2020wum} (dark red points) and the 9.5-year NST dataset~\cite{Abbasi:2021qfz} (olive green points), the IceCube Glashow resonance event~\cite{IceCube:2021rpz} (dark green point),  IceCube-EHE dataset~\cite{IceCube:2018fhm} (black dotted line), the 12.6-year IceCube dataset~\cite{IceCube:2025ezc} (black dashed line), and the upper limit from Auger~\cite{AbdulHalim:2023SN} (navy dashed line).  The theoretical curves demonstrate how PBH-produced sterile neutrinos can generate ultra-high-energy neutrino fluxes.

Remarkably, as shown in the figure, for a suitable choice of parameters, our scenario naturally accounts for the recently reported KM3-230213A event~\cite{KM3NeT:2025npi}, marked by the gray data points. The predicted flux not only matches the observed event energy but also respects the existing constraints from IceCube and Auger, as indicated by the comparison curves. This agreement demonstrates that the heavy sterile neutrino decay channel, seeded by PBH evaporation in the early Universe, provides a viable and predictive explanation for the anomalous neutrino event reported by KM3NeT. 

\begin{figure}[htbp]
    \centering
     \includegraphics[width=0.495\linewidth]{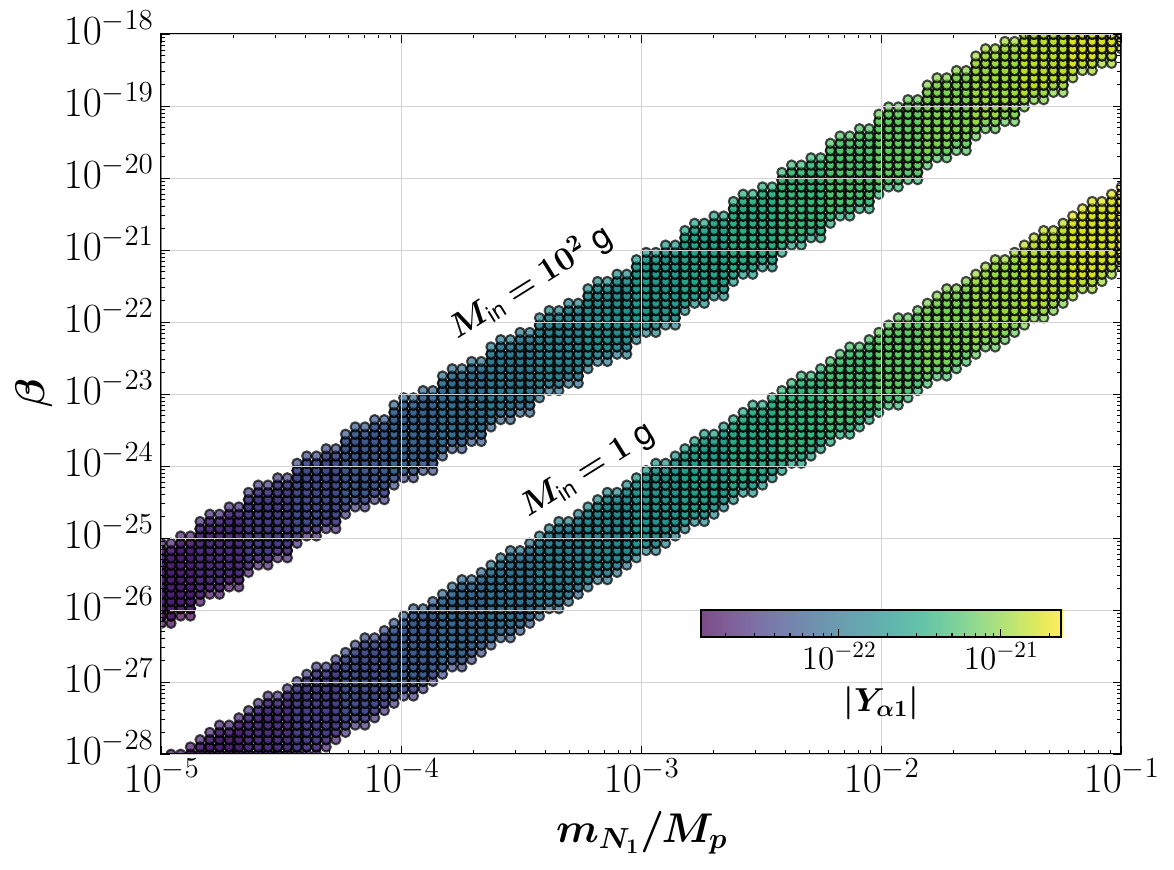}
     \includegraphics[width=0.495\linewidth]{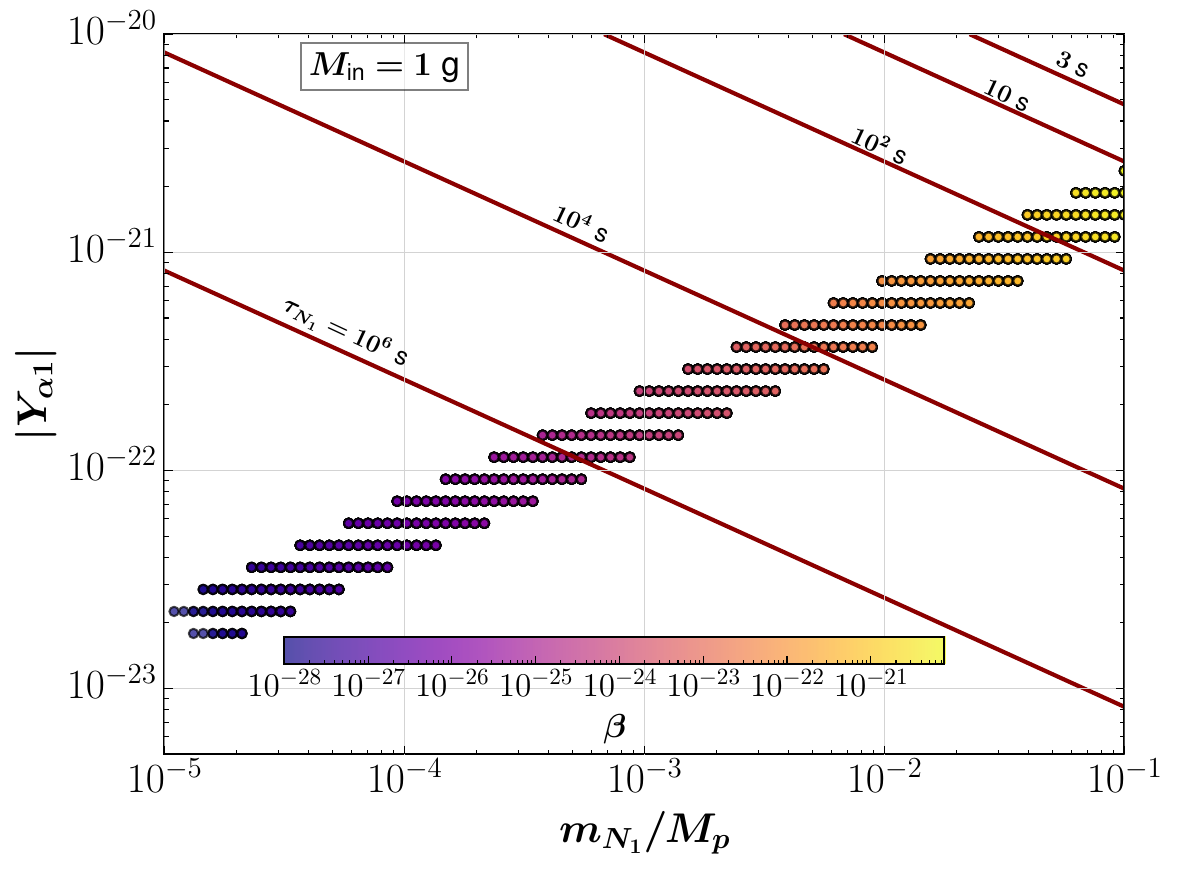}
    \caption{ [Left]: The allowed parameter space in the $\beta$ versus $m_{N_1}/M_p$ plane for two different initial primordial black hole masses: $M_{\rm{in}} = 1~\rm{g}$ and $M_{\rm{in}} = 10^2~\rm{g}$ where the color gradient represents the corresponding Yukawa coupling $|Y_{\alpha 1}|$ values. [Right]: The parameter space in the $|Y_{\alpha 1}|$ versus $m_{N_1}/M_p$ plane for $M_{\rm{in}} = 1~\rm{g}$, with the color gradient indicating the $\beta$ values and the red solid lines represent contours of constant sterile neutrino lifetime $\tau_{N_1}$. 
    }
    \label{fig:param_scan}
\end{figure}

 To determine the parameter space capable of generating the required flux to explain the KM3NeT event, we conduct a random scan over the parameters within the ranges $m_{N_1}\in [10^{-5},10^{-1}]M_p$, $\beta \in [10^{-28},10^{-18}]$, and the Yukawa coupling $|Y_{\alpha 1}|\in [10^{-28},10^{-18}]$, while keeping the initial PBH mass fixed at $M_{\rm in} = 1$ g and $M_{\rm in} = 10^2$ g. {We then extract parameter combinations that produce a spectral peak around the KM3-230213A event energy while ensuring the flux remains consistent with constraints from IceCube and Auger,
purposely selecting the parameter space in our numerical scan to ensure compatibility across experiments. Notably, the full $3\sigma$ confidence interval of the KM3‑230213A flux extends more than an order of magnitude below the IceCube and Auger sensitivities. Hence, if the true flux lies towards the lower end of KM3NeT’s uncertainty band, our scenario remains consistent with the non-observations at IceCube and Auger. However, attempting to fit the central value pushes the predicted signal into the IceCube sensitive regime, recreating the $\sim3\sigma$ tension~\cite{Li:2025tqf}.} 
The results are presented in the left panel of Fig.~\ref{fig:param_scan} in the $\beta$ vs $m_{N_1}/M_p$ plane with the color gradient depicting the value of $|Y_{\alpha 1}|$ for different $M_{\rm in}$ values. 
From the figure, we observe that a decrease in $\beta$ necessitates a corresponding decrease in the sterile neutrino mass $m_{N_1}$ to generate the required flux. This behavior arises because the number density of sterile neutrinos depends on both $\beta$ and $m_{N_1}$, as evident from Eq.~({\ref{eq:nchiatev}}). Similarly, an increase in $m_{N_1}$ requires a larger Yukawa coupling to ensure that the sterile neutrino's lifetime is appropriate for producing ultra-high-energy neutrinos after neutrino decoupling. These neutrinos then redshift to the present day, potentially leading to the observed event.
Thus, it is important to note that while the Yukawa coupling and sterile neutrino mass primarily control the peak position of the spectrum, $\beta$ and $M_{\rm in}$ regulate the peak flux of the spectrum for a given combination of Yukawa coupling and sterile neutrino mass. 
Additionally, as the PBH mass increases, the initial number density of PBHs, $n_{\rm{BH},i}$, decreases, which in turn lowers the sterile neutrino number density, $n_{N_1}$. Consequently, to maintain the observed flux peak, $\beta$ must be increased correspondingly. 
The right panel maps the $|Y_{\alpha 1}|$ vs $m_{N_1}/M_p$ parameter space, with the colorbar illustrating the initial abundance of PBH, $\beta$. We also showcase different contours for different values of $\tau_{N_1}$ as shown by the red solid lines. The viable range of $\tau_{N_1}$ is constrained by two requirements: the lower bound ensures that $N_1$ decay occur after neutrino decoupling, while the upper bound safeguards the successful predictions of BBN light-element abundances~\cite{Kawasaki:2017bqm,Yeh:2024ors}.

\section{Gravitational Wave signatures}\label{section4}
In this section, we discuss two distinct sources of stochastic gravitational wave backgrounds that naturally arise in our scenario. Both GW signals are intrinsically connected to the dynamics responsible for the production of ultra-high-energy neutrinos, thereby offering complementary observational avenues to probe our framework. The first contribution originates from the Hawking evaporation of PBHs, which produces a GW background due to direct graviton emission. The second component arises from graviton Bremsstrahlung emitted during the decay of the heavy sterile neutrinos, whose decay products include the high-energy neutrinos relevant for the KM3NeT event. Importantly, the parameters governing the GW spectra, such as the initial PBH mass, $\beta$, sterile neutrino mass $ m_{N_1} $, and Yukawa coupling $ |Y_{\alpha 1}| $ are the same as those determining the neutrino flux, thereby providing a correlated multimessenger signature of our setup.

\subsection{Gravitational Wave from Hawking Evaporation}\label{section4-1}
The evaporation of PBHs via Hawking radiation results in the emission of gravitons, contributing to a stochastic GW background, with spectral features directly tied to PBH parameters governing the KM3NeT neutrino flux.
The total graviton emission rate per unit time and per unit energy from an ensemble of PBHs is given by:
\begin{equation}
    \frac{d^2\rho_{\rm GW}}{dt dE} \simeq n_{\rm BH} (t) \frac{d^2u_{\rm GW}}{dt dE},
\end{equation}
where \( n_{\rm BH}(t) \) represents the PBH number density at time \( t \), and \( \frac{d^2 u_{\rm GW}}{dt \, dE} \) corresponds to the spectral energy emission rate of gravitons from a single PBH.

Using Eq.~(\ref{eq:nbhconst}) and Eq.~(\ref{eq:PBHmass}), the energy density for gravitational waves at the time of PBH evaporation can be estimated as~\cite{Choi:2024acs}:
\begin{equation}
    \frac{d \rho_{\rm GW, ev}}{d \ln \omega_{\rm ev}} = \frac{27}{64 \pi^3} \frac{M_{\rm in}^2}{M_p^4} n_{\rm BH}(t_i) \omega_{\rm ev}^4 \int_{t_i}^{t_{\rm ev}=t_i+\tau} dt \left(1-\frac{t-t_i}{\tau}\right)^{2/3} \frac{(a_i/a)^3}{e^{\omega_{\rm ev}a_{\rm ev}/a T_{\rm BH}}-1}.
\end{equation}
The present-day GW energy density, $ \rho_{\rm GW,0} $, is connected to the energy density at the time of PBH evaporation, $ \rho_{\rm GW,ev} $, via the redshift relation:
\begin{equation}
    \frac{d \rho_{\rm GW,0}}{d \ln \omega_0} = \frac{d \rho_{\rm GW,ev}}{d \ln \omega_{\rm ev}} \left( \frac{a_{\rm ev}}{a_0} \right)^4,
\end{equation}
where $ \omega_0 $ and $ \omega_{\rm ev} $ are the GW frequencies measured today and at the evaporation epoch, respectively, and $a_{\rm ev}/a_0$ denotes the scale factor ratio between the evaporation time and today.
Since the GW energy density and frequency redshift as $ \rho_{\rm GW} \propto a^{-4}$ and $ \omega \propto a^{-1} $, and by normalizing the present scale factor to unity ($ a_0 = 1 $), the frequency today relates to the evaporation frequency as $\omega_0 = \omega_{\rm ev} \, a_{\rm ev} $.
This leads to the current GW energy density~\cite{Choi:2024acs}:
\begin{eqnarray}\label{eq:rhoGW0}
    \frac{d\rho_{\rm GW,0}}{d \ln \omega_0} &=& \frac{27 g_i}{64 \pi^3} \frac{M_{\rm in}^2}{M_p^4} n_{\rm BH}(t_i) \omega_{0}^4 \int_{t_i}^{t_{\rm ev}} dt \left(1-\frac{t-t_i}{\tau}\right)^{2/3} \frac{(a_i/a)^3}{e^{\omega_{0}/a T_{\rm BH}}-1} \nonumber \\
    &\simeq& 0.5 \beta \frac{M_p^2}{M_{\rm in} } \omega_0^4  \int_{t_i}^{t_{\rm ev}} dt \left(1-\frac{t-t_i}{\tau}\right)^{2/3} \frac{(a_{i}/a)^3}{e^{\omega_{0}/a T_{\rm BH}}-1},
\end{eqnarray}
where we take the initial number density, given by Eq.~(\ref{eq:nBHin}), in the last equality. 
The redshift factor from PBH formation until today can be approximated as:
\dis{
\frac{a_i}{a_0} \simeq \frac{T_{0}}{T_{\rm in}} \left(\frac{g_{*,s}(T_{0})}{g_{*,s}(T_{\rm in})}\right)^{1/3} \simeq 1.8 \times 10^{-29} \left(\frac{M_{\rm{in}}}{1 \ \rm{g}}\right)^{1/6},
}
where we take $T_0 \simeq 2.34 \times 10^{-13} \ \rm{GeV}$, $g_{*,s}(T_0)=3.91$ and $g_*(T_\text{ev}) \simeq g_{*,s}(T_\text{in}) \simeq 100$.  
Similarly, the ratio of the scale factor at evaporation to the present value is approximately:
\begin{equation}\label{eq:aevtoa0}
    \frac{a_{\rm ev}}{a_0} \simeq \frac{T_0}{T_{\rm ev}} \left( \frac{g_{*,s}(T_0)}{g_{*,s}(T_{\rm ev})} \right)^{1/3}
    \simeq 2.3 \times 10^{-24} \left( \frac{M_{\rm in}}{1\, \mathrm{g}} \right)^{3/2}.
\end{equation}
By substituting these relations into Eq.~\eqref{eq:rhoGW0}, the GW energy density spectrum at present can be expressed as:
\begin{eqnarray}
     \frac{d\rho_{\rm GW,0}}{d \ln \omega_0} 
     &\simeq& 3 \times 10^{-74} \beta \ \text{GeV} \left(\frac{M_p}{M_{\rm in}}\right)^{1/2}  \omega_0^4 \ I(\omega_0), 
\end{eqnarray}
where integral $I(\omega_0)$ encapsulates the frequency dependent evolution and is defined by:
\begin{equation} \label{eq:Iw0}
    I(\omega_0) = \int_{t_i}^{t_{\rm ev}} dt \left(1-\frac{t-t_i}{\tau}\right)^{2/3} \frac{a^{-3}}{e^{\omega_{0}/a T_{\rm BH}}-1}.
\end{equation}
Thus the present-day relic abundance of gravitons originating from the direct evaporation of PBHs can be evaluated as:
\begin{eqnarray}\label{eq:omGW_PBHevap}
     h^2 \Omega_{\rm GW} &=& \frac{1}{\rho_{\rm cr,0} h^{-2}} \frac{d\rho_{\rm GW,0}}{d \ln \omega_0}      \simeq 4 \times 10^{-49}  \ \text{GeV}^{-3} \left(\frac{\beta}{10^{-21}}\right) \left(\frac{1 \ \rm g }{M_{\rm in}}\right)^{1/2} \omega_0^4  \ I(\omega_0),
\end{eqnarray}
the present critical energy density is given by $\rho_{\rm cr,0} \simeq 8 h^2 \times 10^{-47} \ \rm{GeV}^4$ with $h \simeq 0.7$.
The peak frequency of the resulting gravitational wave spectrum at present can be analytically determined by extremizing $h^2 \Omega_{\rm GW}$ with respect to $\omega_0$. Under the blackbody approximation for Schwarzschild black holes, the peak frequency is approximately given by $\omega_{\rm peak} \simeq 2.8~ a_{\rm ev} T_{\rm BH}$, which translates to: 
\begin{equation} f_{\rm peak} \simeq 1.6 \times 10^{13} \ {\rm Hz} \left(\frac{M_{\rm in}}{1 \ \rm g}\right)^{1/2}. 
\end{equation}
\begin{figure}[tbp]
    \centering
    \includegraphics[width=0.7\textwidth]{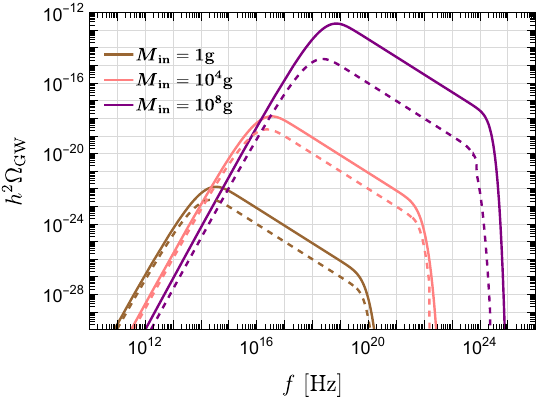}
    \caption{Gravitational wave spectrum from direct evaporation of PBHs for different initial masses, $M_{\rm in} = 1 \ \rm{g}$, $10^4 \ \rm{g}$, and $10^8 \ \rm{g}$. The solid lines correspond to the full numerical evaluation including greybody factors, whereas the dashed lines show the analytical approximation as in Eq.~(\ref{eq:omGW_PBHevap}). Here, we have fixed $\beta = 10^{-21}$.}
    \label{fig:GW_direct}
\end{figure}
During the radiation-dominated epoch, the scale factor evolves with time as $a = A t^{1/2}$, where the coefficient $A$ is determined by $a = a_{\rm ev} \left( \frac{t}{\tau_{\rm BH}}\right)^{1/2} = A t^{1/2}$. Using Eqs.~(\ref{eq:lifetime}) and (\ref{eq:aevtoa0}), the coefficient $A$ is explicitly given by: \begin{eqnarray}\label{eq:Aquant} A = a_{\rm ev} \left(\frac{1}{\tau}\right)^{1/2} \simeq 7.6 \times 10^{-32} \ M_{p}^{1/2}. \end{eqnarray}
Accordingly, the integral in Eq.~(\ref{eq:Iw0}) can be expressed as:
\begin{eqnarray}
     I(\omega_0) &=& A^{-3} \int_{t_i}^{t_{\rm ev}} dt \left(1-\frac{t-t_i}{\tau}\right)^{2/3} \frac{t^{-3/2}}{\exp\left[\alpha t^{-1/2}\left(1-\frac{t-t_i}{\tau}\right)^{1/3}\right]-1},
\end{eqnarray}
where $\alpha\equiv\frac{\omega_0 M_{\rm in}}{A M_p^2}$. 
This integral, which carries an explicit dependence on $\omega_0$, is typically evaluated numerically for each frequency. The physical frequency can be related as $f = \omega_0 / 2\pi$.
For this calculation, the PBH formation time $t_i$ is determined by Eq.~(\ref{eq:MassPBH}), along with the energy density at formation $\rho(T_{\rm in}) = 3 M_p^2 \mathcal{H}^2(T_{\rm in})$, resulting in 
\begin{equation} t_i = \frac{M_{\rm in}}{8\pi \gamma M^2_p}. 
\end{equation} The evaporation time is then approximately given by $t_{\rm ev} \simeq \tau$, since $t_i \ll \tau$.

The resulting gravitational wave spectrum from direct PBH evaporation during the radiation-dominated era is displayed in Fig.~\ref{fig:GW_direct} as a function of frequency $f$, for three benchmark values of the initial PBH mass: $M_{\rm in} = 1 \ \rm{g}$, $10^4 \ \rm{g}$, and $10^8 \ \rm{g}$, with a fixed value of $\beta = 10^{-21}$. The solid lines correspond to the full numerical computation that incorporates greybody factors, whereas the dashed lines represent the analytical approximation based on the geometric optics limit. It is worth highlighting that as the PBH mass increases, both the peak frequency and amplitude of the gravitational wave spectrum increase, shifting the spectrum towards higher frequencies. This behavior arises because heavier PBHs form at later times as well as evaporate for a longer period of time, resulting in a reduced cosmological redshift of the emitted gravitons and thus leading to a higher observed frequency and amplitude today.

\subsection{Graviton Bremsstrahlung in Sterile Neutrino Decay}\label{section4-2}
We now examine the gravitational wave signal arising from graviton bremsstrahlung during sterile neutrino decays. While all particle decays inherently produce gravitational radiation through graviton emission, this process is typically suppressed by the Planck-scale coupling ($\sim M_p^{-1}$) between matter fields and gravitons. The branching ratio for graviton emission scales as $Br(N_1\to \nu+ h +{\rm graviton}) \sim (M_{N_1}/M_p)^2$, making it significant only for superheavy particles~\cite{Choi:2024acs,Hu:2024awd}.
This suppression is overcome in our scenario, where the sterile neutrinos possess masses near the Planck scale ($M_{N_1} \sim M_p$), precisely the regime required to generate the ultra-high-energy neutrinos observed by KM3NeT through decays $N_1 \to \nu h$. The resulting graviton spectrum carries unique imprints of both the sterile neutrino mass and its decay kinematics, providing a complementary signature to the Hawking evaporation signal discussed in the previous section.

\begin{figure}[htbp]
    \centering
    \includegraphics[width=0.7\textwidth]{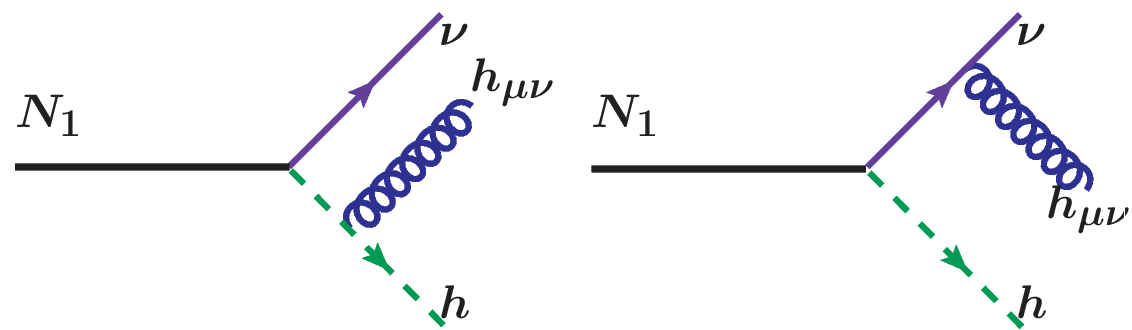}
    \caption{Feynman diagrams representing the decay of sterile neutrino $N_1$ into neutrino ($\nu$) and SM Higgs $h$ along with a graviton ($h_{\mu\nu}$) bremsstrahlung. }
    \label{fig:graviton_brems}
\end{figure}

The energy density of gravitons $\rho_{\rm GW}$ produced through bremsstrahlung in sterile neutrino decays evolves according to the Boltzmann equation:
\begin{equation}
    \frac{d}{dt} \left(\frac{d \rho_{\rm GW}}{dE_{\rm GW}}\right) + 4 \mathcal{H} \frac{d \rho_{\rm GW}}{dE_{\rm GW}} = n_{N_1}(a_{\rm ev}) \left(\frac{a_{\rm ev}}{a}\right)^3 \frac{d \Gamma_{N_1 \rightarrow \rm GW}}{dE_{\rm GW}} E_{\rm GW},
\end{equation}
which can be rewritten in terms of the scale factor 
$a$ as:
\begin{eqnarray}\label{eq:bremsboltzeq}
    \frac{d}{da} \left(a^4 \frac{d \rho_{\rm GW}}{d \ln E_{\rm GW}}\right) = \frac{n_{N_1}(a_{\rm ev}) a_{\rm ev}^3}{\mathcal{H}} \frac{d \Gamma_{N_1 \rightarrow \rm GW}}{dE_{\rm GW}} E_{\rm GW}^2.
\end{eqnarray}
For the non-relativistic sterile neutrino decays at the rest frame, the differential decay rate can be estimated as:
\dis{
\frac{d \Gamma_{N_1 \rightarrow \rm GW}}{dE_{\rm GW}} = \frac{|Y_{\alpha_1}|^2 m_{N_1}^3 }{64 \pi^3 M_p^2 E_{\rm GW}} \mathcal{G}(E_{\rm GW}/m_{N_1}),
}
where the form factor: $\mathcal{G}(x) = (x-1)^2(1-2x)$
 encodes the graviton energy dependence. Integrating Eq.~(\ref{eq:bremsboltzeq}),
from the period of complete PBH evaporation ($a_{\rm ev}$) to the sterile neutrino decay ($a_{N_1}$) yields:
\begin{eqnarray}
     a_{N_1}^4 \frac{d \rho_{\rm GW}(a_{N_1})}{d \ln E_{\rm GW}} &=&  \frac{|Y_{\alpha 1}|^2 m_{N_1}^3}{64 \pi^3 M_p^2} E_{\rm GW} \mathcal{G}(E_{\rm GW}/m_{N_1}) n_{N_1}(a_{\rm ev}) a_{\rm ev}^3 \int_{a_{\rm ev}}^{a_{N_1}} da \frac{1}{\mathcal{H}} \nonumber \\
     &=&  \frac{|Y_{\alpha 1}|^2 m_{N_1}^3}{64 \pi^3 M_p^2} E_{\rm GW} \mathcal{G}(E_{\rm GW}/m_{N_1}) n_{N_1}(a_{\rm ev}) a_{\rm ev}^3 ~\frac{2}{3 A^2} (a_{N_1}^3 - a_{\rm ev}^3),
\end{eqnarray}
where we account for radiation-dominated expansion and the constant $A$ is given in Eq.~(\ref{eq:Aquant}).
Now using the comoving number density of the sterile neutrino just after PBH evaporation given in Eqs.~(\ref{eq:nchiatev}), we obtain the GW energy density at $a_{N_1}$:
\begin{eqnarray}
    \frac{d \rho_{\rm GW}(a_{N_1})}{d \ln E_{\rm GW}} &\simeq&  5 \times 10^{-6} ~\beta~ |Y_{\alpha 1}|^2 \frac{m_{N_1} M_p^5}{M_{\rm in}^3} E_{\rm GW} \mathcal{G}(E_{\rm GW}/m_{N_1}) \bfrac{a_{\rm ev}}{a_{N_1}}\left[1-\left(\frac{a_{\rm ev}}{a_{N_1}}\right)^3\right]. 
\end{eqnarray}
The present-day GW relic abundance is then:
\begin{equation}
    h^2 \Omega_{\rm GW} = \frac{1}{\rho_{\rm cr, 0} h^{-2}} \frac{d \rho_{\rm GW} (a_{N_1})}{d \ln E_{\rm GW}} \left(\frac{a_{N_1}}{a_0}\right)^4, 
\end{equation}
with $E_{\rm GW} = 2\pi f (a_0/a_{N_1})$. Therefore, $h^2 \Omega_{\rm GW}$ can be written by:
\begin{eqnarray} \label{eq:finOmGW}
     h^2 \Omega_{\rm GW} &\simeq& \frac{5 \times 10^{-6}}{\rho_{\rm cr, 0} h^{-2}} 
     \beta ~ |Y_{\alpha 1}|^2 \frac{m_{N_1} M_p^5}{M_{\rm in}^3} E_{\rm GW}\mathcal{G}(E_{\rm GW}/m_{N_1}) \bfrac{a_{\rm ev}}{a_{N_1}}\left[1-\left(\frac{a_{\rm ev}}{a_{N_1}}\right)^3\right]
     \left(\frac{a_{N_1}}{a_0}\right)^4 \nonumber \\
     &\simeq& 4 \times 10^{41} \ {\rm{GeV}^{-4}}~ \beta ~ |Y_{\alpha 1}|^2 f \frac{m_{N_1} M_p^5}{M_{\rm in}^3} \mathcal{G}(E_{\rm GW}/m_{N_1}) \bfrac{a_{\rm ev} }{a_0}\left(\frac{a_{N_1}}{a_0}\right)^2.
\end{eqnarray}
Substituting the scale factor relations from Eqs.~(\ref{eq:aevtoa0}) and (\ref{eq:axtoa0}) gives:
\begin{eqnarray}\label{eq:omGW_N}
     h^2 \Omega_{\rm GW} &\simeq& 1.5 \times 10^{-18} \bfrac{\beta}{10^{-21}} \left(\frac{f}{10^{31} \ \rm Hz}\right) \left(\frac{1 \ \rm{g}}{M_{\rm in}}\right)^{3/2} \mathcal{G}(E_{\rm GW}/m_{N_1}) .
\end{eqnarray}
\begin{figure}[tbp]
    \centering
    \includegraphics[width=0.495\textwidth]{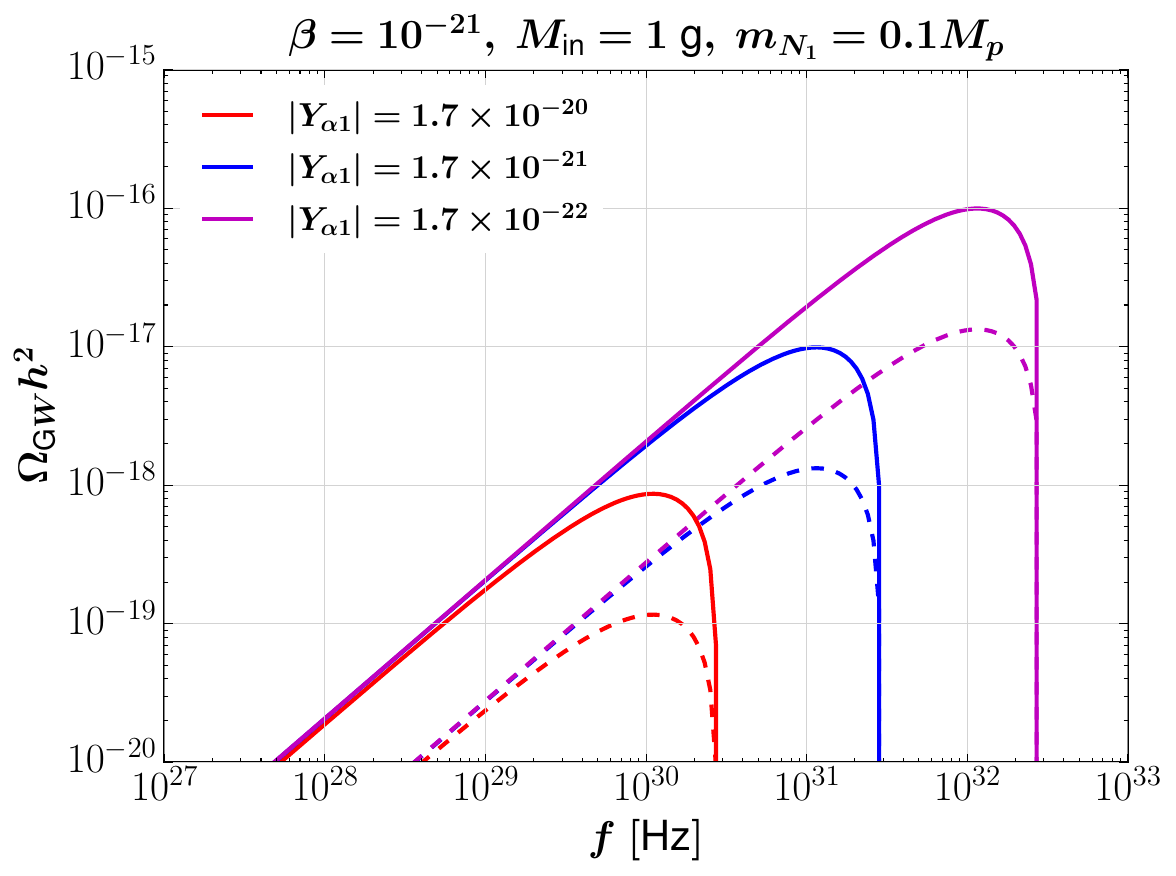}
    \includegraphics[width=0.495\textwidth]{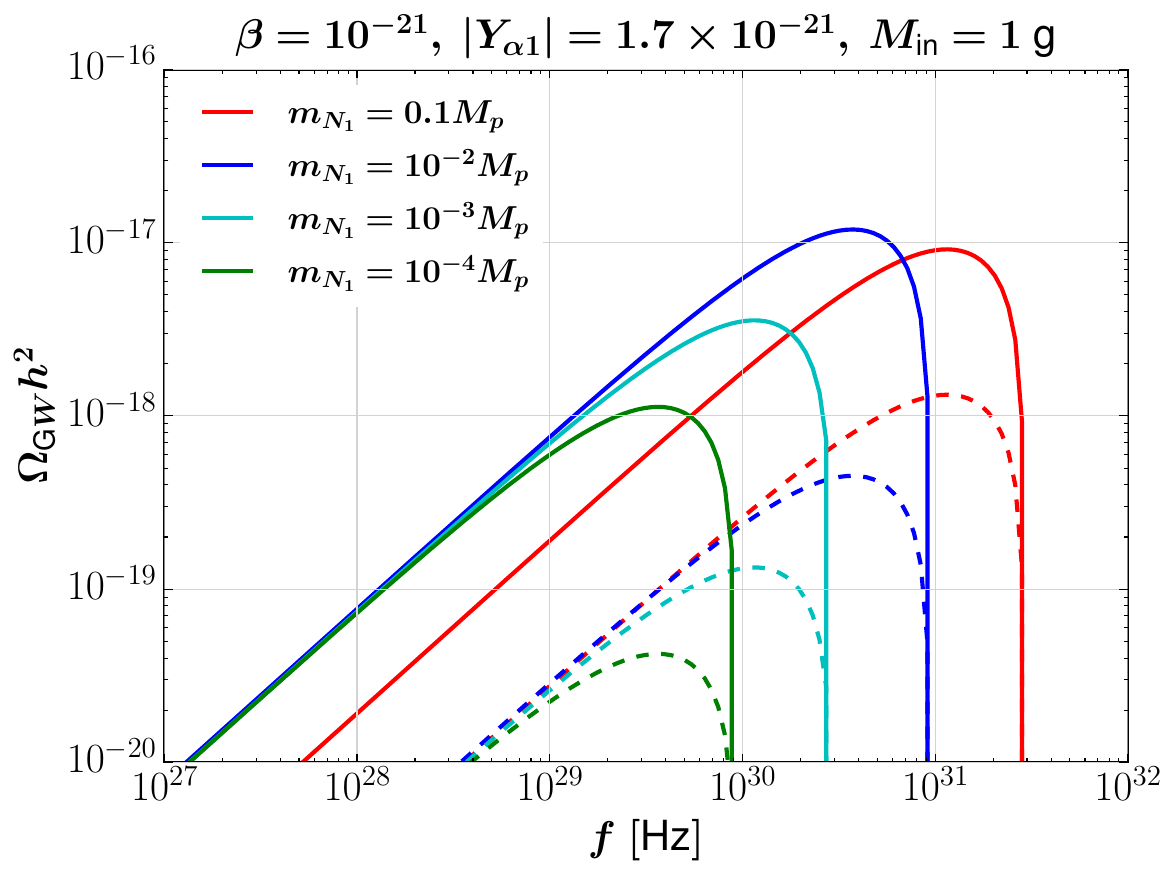}
    \caption{Gravitational wave spectra for graviton bremsstrahlung from sterile neutrino decays for $m_{N_1}= 0.1 M_p$ (left) and the fixed Yukawa coupling $|Y_{\alpha 1}|=1.7 \times 10^{-21}$ (right). Solid curves show full numerical results, while dashed curves correspond to the analytic estimation (Eq.~\ref{eq:omGW_N}). Different colors represent different values for the corresponding the coupling $|Y_{\alpha 1}|$ and mass $m_{N_1}$. }
    \label{fig:OmGWfromN}
\end{figure}
As the GWs are produced from the decay of sterile neutrino, their energy at the time of production is bounded by $E_{\rm GW} \leq m_{N_1}/2$. Accordingly, the peak frequency of the resulting GW spectrum is given by:
\begin{eqnarray}
    f_{\rm peak} = \frac{m_{N_1}}{4 \pi} \left(\frac{a_{ N_1}}{a_0}\right) 
    \simeq {5.12 \times 10^{31} \ \text{Hz} \left(\frac{10^{-21}}{|Y_{\alpha 1}|}\right) \left(\frac{m_{N_1}}{0.1 M_p}\right)^{1/2}}
\end{eqnarray}

Fig.~\ref{fig:OmGWfromN} presents the gravitational wave spectrum as a function of frequency, showing two key parameters: the Yukawa coupling $|Y_{\alpha 1}|$ and the sterile neutrino mass $m_{N_1}$ dependencies. The important features that emerge from the analysis are that the spectrum remains invariant under parameter changes for $E_{\rm GW}\lesssim m_{N_1}$ as predicted by Eq.~(\ref{eq:omGW_N}). Heavier particles produce higher peak frequencies as a consequence of greater energy transfer to gravitons. Smaller Yukawa coupling values delay the decay epoch, reducing cosmological redshift and thereby increasing the peak frequency. The agreement between numerical results (solid curves) and analytical estimations (dashed curves) validates our treatment of the bremsstrahlung process.

Fig.~\ref{hc_plot} shows the dimensionless characteristic strain $h_c$, a key observable for GW detection, as a function of gravitational wave frequency $f$. The dimensionless characteristic strain relates to the GW relic density as: \begin{equation}
    h_c = f^{-1}\sqrt{\frac{3 \mathcal{H}_0^2}{4 \pi^2} \Omega_{\rm GW} }  \simeq 8.93 \times 10^{-19} \sqrt{\Omega_{\rm GW}h^2} \left(\frac{ \rm Hz}{f}\right),
\end{equation}
where the Hubble rate at present $H_0 = 100 h \ \text{km} \ \text{s}^{-1} \ \text{Mpc}^{-1} \simeq 3.24 \times 10^{-18} h \ \text{s}^{-1}$.
The plot compares the GW spectra from the two distinct sorces: direct emission of GWs via PBH evaporation with different initial masses, $M_{\rm in} = 1 \ \rm{g}, 10^{4} \ \rm{g}$ and $10^8 \ \rm{g}$ and the Graviton bremsstrahlung from sterile neutrino decays for various Yukawa couplings, as mentioned in the inset of the figure, with $M_{\rm in} = 1 \ \rm{g}$. We further overlay the sensitivity curves for the current and prospective experimental setups that are capable of probing high-frequency gravitational waves such as optically levitated sensors, enhanced magnetic conversion (EMC) experiments, and the inverse Gertsenshtein effect. Representative experiments include JURA, ALPS II, OSQAR, IAXO, and CAST, whose projected sensitivities are adopted from Ref.~\cite{Aggarwal:2020olq}. The gray shaded region corresponds to the reach of resonant cavity experiments~\cite{Herman:2022fau}.

The GW energy density evolves as $\rho_{\rm GW} \propto a^{-4}$, contributing to the effective neutrino species: 
\begin{equation}
\rho_{\rm rad} = \rho_\gamma \left(1+\frac{7}{8}\left(\frac{4}{11}\right)^{4/3} N_{\rm eff}\right),
\end{equation}
where $N_{\rm eff} = N_{\rm eff}^{\rm SM} + \Delta N_{\rm eff}$ with $N_{\rm eff}^{\rm SM}=3.046$~\cite{Mangano:2005cc}. The GW contribution is:
\begin{equation}
\Delta N_{\rm eff} = \frac{8}{7}\left(\frac{11}{4}\right)^{4/3} \frac{\rho_{\rm GW}}{\rho_\gamma}=\frac{120}{7 \pi^2} \left(\frac{11}{4}\right)^{4/3} \frac{\rho_{\rm cr, 0}}{T_0^4} \Omega_{\rm GW}^{\rm max}.
\end{equation}
Thus current Planck constraints ($\Delta N_{\text{eff}} < 0.30$ at 95$\%$ C.L.)~\cite{Planck:2018vyg} as well as the  future sensitivities from CMB-S4~\cite{CMB-S4:2016ple}, Euclid~\cite{laureijs2011euclid}, and CMB-CVL~\cite{Ben-Dayan:2019gll} are shown as dark yellow bounds in Fig.~\ref{hc_plot}. It is evident that the GW backgrounds originating from both PBH evaporation and sterile neutrino bremsstrahlung decay are consistent with the existing limits on $\Delta N_{\rm eff}$. 
{The stochastic GW background resulting from direct graviton emission from PBH with mass $M_{\rm} = 10^8 \ \rm{g}$ may fall within the detection threshold of the future CMB sensitivies.}
However, future CMB measurements with improved sensitivities, along with dedicated high-frequency GW detection experiments, may have the potential to probe this scenario, offering a complementary avenue to explore such high-frequency GW signals.
\begin{figure}[tbp]
\centering
\includegraphics[width=0.7\textwidth]{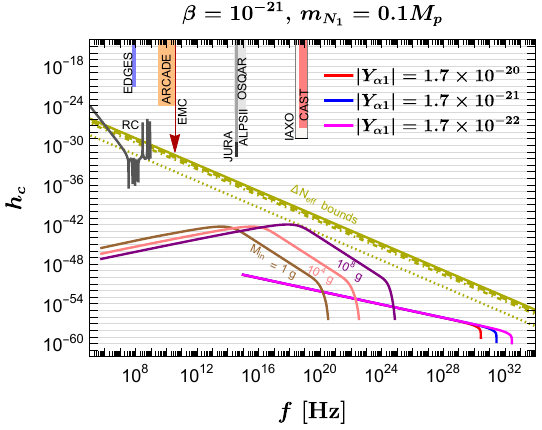}
\caption{The characteristic strain $h_c$ as a function of frequency, illustrating the gravitational wave signals arising from both PBH evaporation with different initial mass of PBHs and sterile neutrino bremsstrahlung processes with $M_{\rm in} = 1 \ \rm{g}$. The shaded regions indicate the current and projected sensitivities of various high-frequency GW detection techniques. The yellow contours correspond to existing and future constraints on $\Delta N_{\rm eff}$. See text for further details.}
\label{hc_plot}
\end{figure}

{In addition to constraints on $\Delta N_{\rm eff}$, BBN also imposes stringent limits on the abundance and lifetime of the sterile neutrino as its decay occurs after neutrino decoupling and must not disrupt the observed light-element abundances. According to the analysis in ~\cite{Kawasaki:2017bqm}, BBN constrains the product  $m_X \frac{n_X}{s}$ (where $X$ denotes the long-lived decaying particle) to avoid excessive energy injection into the primordial plasma from hadronic or electromagnetic decay products.
 In our scenario, the heavy sterile neutrino decays predominantly via $N_1 \to \nu h$, $N_1 \to \nu Z$ and $N_1 \to \ell^{\pm} W^{\mp}.$ Although these channels ultimately produce secondary hadrons and photons, the overall energy injection is highly suppressed due to the extremely small comoving yield of the sterile neutrinos produced via PBH evaporation. Consequently, even with a sterile neutrino mass near the Planck scale, the ratio of $N_1$ energy density to radiation energy density $(\rho_{N_1}/\rho_r)$  around the epoch of its decay  is negligible. For a {benchmark scenario with $m_{N_1}=0.1 M_p, \beta=10^{-21}, |Y_{\alpha 1}|=10^{-21}$, we estimate the energy density ratio $(\rho_{N_1}/\rho_r)$ to be of the order $\mathcal{O}(10^{-7})$ at $a_{N_1}$, which leads to $m_{N_1} \frac{n_{N_1}}{s}|_{a_{N_1}}= \frac{3}{4} T_{N_1} (\rho_{N_1}/\rho_r)|_{a_{N_1}} \sim 10^{-12}$ GeV.}
 Thus the sterile neutrino decay lifetime in our model can conservatively extend up to $\tau_{N_1} \sim 10^6$ s without conflicting with BBN constraints~\cite{Kawasaki:2017bqm,Yeh:2024ors}.
}

\section{Leptogenesis}\label{section5}

{As is well known, the Majorana nature of the heavy RHNs inherently leads to lepton number violation. The lepton asymmetry is dynamically produced from the CP-violating decays of the
heavy Majorana neutrinos~\cite{Buchmuller:2004nz,Buchmuller:2002rq,Plumacher:1996kc,Giudice:2003jh,Barbieri:1999ma}. There are also washout processes present which compete with
these decays and act to reduce the overall asymmetry. The final lepton asymmetry is partially reprocessed to a baryon asymmetry via weak sphaleron processes which proceed at
unsuppressed rates above the electroweak scale.
 This lepton asymmetry can arise from the interference between the tree-level and one-loop decay diagrams, as illustrated in Fig.~\ref{fig:leptofeyn}, which include both the self-energy and vertex corrections. 
\begin{figure}[htbp]
    \centering
    \includegraphics[width=0.8\linewidth]{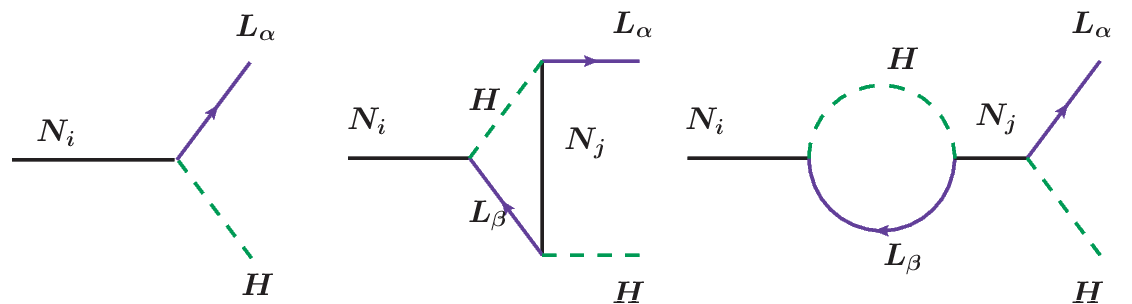}
    \caption{{Feynman diagrams for tree and one-loop level decays
of RHNs giving rise to nonzero CP asymmetry.}}
    \label{fig:leptofeyn}
\end{figure}

The $CP$ asymmetry parameter is defined as,
\begin{equation}    \varepsilon_i=\frac{\Gamma(N_i \rightarrow L H)-\Gamma(N_i \rightarrow \bar{L} \bar{H})}{\Gamma(N_i \rightarrow L H)+\Gamma(N_i \rightarrow \bar{L} \bar{H})}.
\end{equation}
Considering the interference of tree level diagram with one-loop diagrams we have \cite{Davidson:2002qv}:
\begin{equation}
    \varepsilon_i = -\frac{1}{8\pi}\sum_{j\neq i} \frac{Im[(Y^\dagger Y)_{ij}^2]}{(Y^\dagger Y)_{ii}}\bigg[f_v\bigg(\frac{M_j^2}{M_i^2}\bigg)+f_s\bigg(\frac{M_j^2}{M_i^2}\bigg)\bigg]
\end{equation}
where the functions $f_v(x)$ and $f_s(x)$ arise from the one-loop vertex and self-energy corrections, respectively, and are given by:
\begin{equation}
\begin{split}
    &f_s(x)=\frac{\sqrt{x}}{1-x};\\
    &f_v(x)= \sqrt{x}\bigg[1-(1+x)\ln\bigg(\frac{1+x}{x}\bigg)\bigg]
\end{split}
\end{equation}
}

{As discussed in the previous sections, in our setup, the decay of $N_1$ responsible for generating the UHE neutrino flux in the early Universe cannot be used to generate the lepton asymmetry that would eventually be converted to the baryon asymmetry. This is because the decay of $N_1$ occurs after neutrino decoupling, ensuring that the produced high-energy neutrinos do not thermalize again. Consequently, this decay takes place well after the sphaleron transition temperature, which is necessary for the successful conversion of lepton asymmetry into baryon asymmetry. Additionally, the extremely small coupling involved in the decay of $N_1$ results in a very small CP asymmetry parameter, $\varepsilon_1 \lesssim 10^{-30}$, making the lepton asymmetry produced from $N_1$ decay insignificant.

On the other hand, the other two RHNs, $N_2$ and $N_3$, which are introduced to explain the neutrino oscillation data, have significant couplings as discussed in Section~\ref{sec:nucouplings}. These RHNs decay earlier and can thus generate the required amount of lepton asymmetry to explain the observed baryon asymmetry of the Universe. In this section, we focus on leptogenesis driven by $N_2$ and $N_3$, taking into account their decay and inverse decay processes, as well as lepton number violating scattering processes.

Defining the comoving abundance of RHNs as $N_{N_i} = n_{N_i}/n_\gamma$, where $n_{N_i}$ is the number density of $i^{\rm th}$ RHN and $n_\gamma$ is the photon density and  the number density of the $B-L$ asymmetry in a comoving volume as $N_{B-L}=n_{B-L}/n_\gamma$, the Boltzmann equations (BEs) to track the cosmological evolution of $N_{B-L}$ and $N_{N_i}$ are given by~\cite{Buchmuller:2004nz,Buchmuller:2002rq,Plumacher:1996kc,Paul:2025iks}:

\begin{eqnarray}\label{eq:BE}
        \frac{d N_{N_2}}{dz} &=& -  (D_2+S_2)(N_{N_2}-N_{N_2}^{\rm eq}),\nonumber\\
        \frac{d N_{N_3}}{dz} &=& - (D_3+S_3)(N_{N_3}-N_{N_3}^{\rm eq}),\nonumber\\
        \frac{d N_{\rm B-L}}{dz} &=& - \sum_{i=2}^3 [\varepsilon_i D_i(z) (N_{N_i}-N_{N_i}^{\rm eq}) + W_{i}(z)N_{\rm B-L}(z)],\nonumber\\\label{eq:be}
\end{eqnarray}
where $z=m_{N_2}/T$ and  $D_i(z)={\Gamma_{Di}}/{\mathcal{H}z}$ which accounts for the decays of $N_i$. $S_i$, $i=1,2,3$ constitutes $\Delta L=1$ scattering processes involving $N_i$. $W_i$ represents the washout term due to inverse decays, $\Delta L=1$ and $\Delta L =2$ scatterings. As the lightest RHN $N_2$ mass is chosen to be $\lesssim 10^{14}$ GeV, the $\Delta L =2$ washout processes are not efficient~\cite{Buchmuller:2002rq,Plumacher:1996kc} and hence we do not include them in our calculations. As $N_1$ decay neither contribute to the lepton asymmetry, nor does the scatterings involving $N_1$ as a mediator contribute to the washout processes because of the heavy mass of $N_1$ and feeble coupling, so we exclude tracking the evolution of $N_1$, in this analysis.

The final $\rm B-L$ asymmetry just before the electroweak sphaleron freeze-out is converted into the observed baryon to photon ratio by the relation:

\begin{equation}
    \eta_B = \frac{C_{B\to L}}{d}  N^{\rm f}_{\rm B-L} = 1.23 \times 10^{-2} N^{\rm f}_{\rm B-L}
\end{equation}
where $C_{B\to L}= 0.35443$ is the fraction of $\rm B-L$ asymmetry converted into baryon asymmetry by sphaleron process, and $d = g_{*}(T_{N_2})/g_{*}(T_{\rm CMB})=110/3.91=28.133$ is the dilution factor calculated assuming standard photon production from the onset of leptogenesis till recombination. Thus in order to obtain the observed baryon asymmetry $\eta_{B}=6.1\pm0.3 \times 10^{-10}$, the required lepton asymmetry is $[4.60,5.08]\times 10^{-8}$.

\begin{figure}[htbp]
    \centering
    \includegraphics[width=0.6\linewidth]{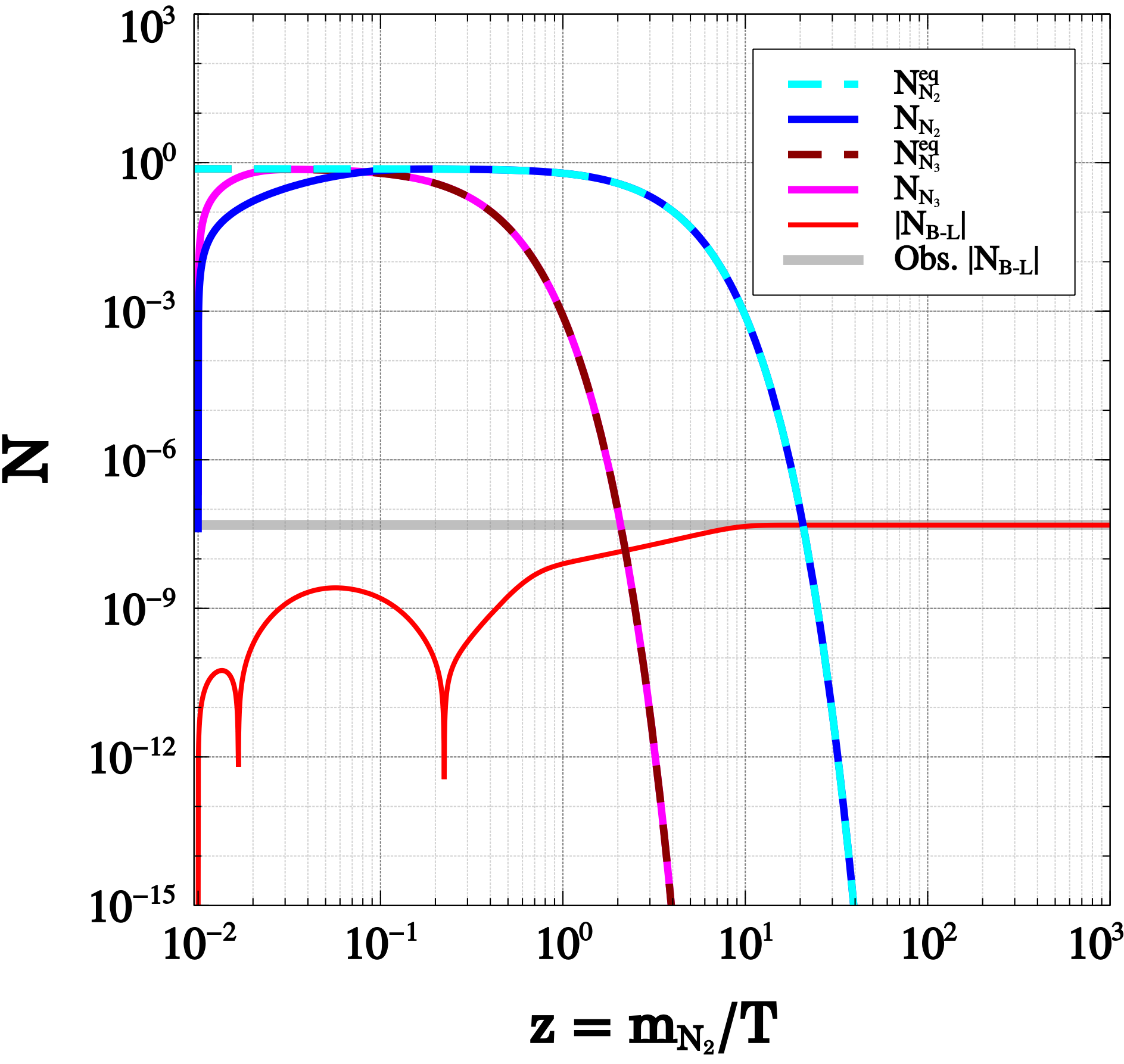}
    \caption{Cosmological evolution of comoving number densities of RHNs $N_2$ and $N_3$ along with $B-L$ asymmetry.}
    \label{fig:leptoevo}
\end{figure}

In Fig.~\ref{fig:leptoevo}, we show the evolution of the comoving number densities of the RHNs $N_2$ (solid blue) and $N_3$ (solid magenta), as well as the generated $B-L$ asymmetry $|N_{\rm B-L}|$ (solid red). The corresponding CP asymmetry parameters for $N_2$ and $N_3$ are found out to be $\varepsilon_2=1.08 \times 10^{-3}$ and $\varepsilon_3= 7.8 \times 10^{-5}$ respectively. 
These are calculated for the benchmark RHN masses given in Eq.~(\ref{eq:rhnMR}) and the Yukawa couplings obtained via the Casas-Ibarra parametrization as given in Eq.~(\ref{eq:yukawa}). The washout parameter for each RHN, defined as
\begin{equation}
    K_i=\frac{\Gamma_i(z=\infty)}{\mathcal{H}(z=1)}
    =\frac{\tilde{m}_i m_{N_i}^2/(8\pi v^2)}{1.66 \sqrt{g_*} m_{N_i}^2/M_{\rm Pl}}
    =\frac{\tilde{m}_i}{1.0697\times 10^{-3}~\rm eV},
\end{equation}
where $\tilde{m}_i=\frac{v^2 (Y^\dagger Y)_{ii}}{2 m_{N_i}}$, determines the washout regime. If $K_i<1$, the $i^{\rm th}$ RHN is in the weak washout regime, while $K_i>1$ indicates the strong washout regime.
For the benchmark parameters considered here, we find $K_2=23.75$ and $K_3=95.63$, indicating that both $N_2$ and $N_3$ are in the strong washout regime. This clearly demonstrates that it is possible to achieve the required baryon asymmetry by leveraging the decay of $N_2$ and $N_3$ while late time decay of  $N_1$ is used for generating the UHE neutrino flux needed to explain the KM3NeT event.

Here, it is worth mentioning that, though we focus on the non-resonant thermal leptogenesis as we maintain a heirarchy between $N_2$ and $N_3$ masses such that $m_{N_3} = 10 ~m_{N_2}$, one can also achieve the required baryon asymmetry via resonant leptogenesis~\cite{Pilaftsis:2003gt} in this setup. We note that the non-thermal contributions to lepton asymmetry from $N_2$ and $N_3$ decays, which are produced from PBH evaporation, are suppressed in our model due to small $\beta$ value required to explain the KM3NeT event as summarized in Fig.~\ref{fig:param_scan},  and thus it does not affect the lepton asymmetry results obtained in Fig.~\ref{fig:leptoevo}. In order to get the observed value for baryon asymmetry only from such non-thermal contributions, one has to consider larger values of $\beta$ ~\cite{JyotiDas:2021shi, Bernal:2022pue, Calabrese:2023key}.   The same decays of $N_{2,3}$ can also be used for generating both the baryon asymmetry via leptogenesis as well as in producing the dark matter relic abundance in asymmetric dark matter or cogenesis setups~\cite{Falkowski:2011xh,Petraki:2013wwa,Zurek:2013wia,Borah:2024wos,Mahapatra:2023dbr}.  
}

\section{Conclusion}\label{section6}
The detection of the ultra-high-energy neutrino event KM3-230213A by the KM3NeT collaboration presents a significant challenge to conventional astrophysical and cosmogenic models. In this work, we proposed a novel cosmological scenario to explain this event, invoking the interplay between primordial black holes and super-heavy sterile neutrinos. Our framework proposes that PBHs formed in the early Universe emit sterile neutrinos via Hawking evaporation, which subsequently decay into active neutrinos. The resulting neutrino flux, after cosmological redshift, aligns with the energy and isotropic origin of the KM3-230213A event.

The model employs a type-I seesaw mechanism, where two sterile neutrinos generate light neutrino masses consistent with oscillation data, while a third, feebly coupled sterile neutrino with a Planck-scale mass decays to produce the observed ultra-high-energy neutrinos. The decay kinematics and redshift evolution yield a flux peaking at $\mathcal{O}(100)$ PeV, matching the KM3NeT observation while respecting constraints from IceCube and the Pierre Auger Observatory. {The sterile neutrinos with significant coupling so as to generate the neutrino masses consistent with oscillation data can also be leveraged to produce the matter-antimatter asymmetry of the Universe via the baryo-lepto-genesis route.}

A distinctive feature of our scenario is the prediction of two complementary gravitational wave signatures. The first arises from gravitons emitted during PBH evaporation, producing a stochastic GW background with a spectrum dependent on the PBH initial mass and abundance. The second originates from graviton Bremsstrahlung during sterile neutrino decays, generating ultra-high-frequency GWs. These GW signals are intrinsically tied to the parameters governing the neutrino flux {\it i.e.} PBH mass, PBH energy density relative to radiation at its formation time , sterile neutrino mass and the Yukawa coupling, offering a multi-messenger probe of the mechanism.

In summary, our work establishes a compelling cosmological origin for the KM3-230213A event, rooted in the dynamics of PBHs and sterile neutrinos. It highlights the synergy between high-energy neutrino astronomy and gravitational wave cosmology, opening new avenues to explore early Universe physics. Future observations of ultra-high-energy neutrinos and high-frequency GWs will critically test this framework, potentially unveiling the role of PBHs and super-heavy particles in shaping the cosmic neutrino and stochastic gravitational wave backgrounds. 

\acknowledgments
The authors acknowledge the financial support from National Research Foundation(NRF) grant
funded by the Korea government (MEST) NRF-2022R1A2C1005050. SM acknowledges Partha Kumar Paul for useful discussion.

\bibliographystyle{JCAP}
\bibliography{reference}
\end{document}